\documentclass[showpacs,prd,aps,nofootinbib]{revtex4}

\pdfoutput=1

\usepackage[colorlinks,urlcolor=blue,citecolor=blue]{hyperref}

\usepackage{amsmath}
\usepackage{amssymb}
\usepackage{latexsym}
\usepackage{amsfonts}
\usepackage{epsfig}
\usepackage{psfrag}
\usepackage{graphicx}

\usepackage{mathtools} 
\def\hhref#1{\href{http://arxiv.org/abs/#1}{arXiv:#1}} 

\usepackage{color}

\newcommand{\bea}{\begin{eqnarray}}
\newcommand{\ea}{\end{eqnarray}}
\newcommand{\eea}{\end{eqnarray}}

\begin{document}

\title{On the Time Dependence of Adiabatic Particle Number}

\author{Robert~Dabrowski and Gerald~V.~Dunne}

\affiliation{Department of Physics, University of Connecticut,
Storrs CT 06269-3046, USA}

\begin{abstract}

We consider quantum field theoretic systems subject to a time-dependent perturbation, and discuss the question of defining a time dependent particle number not just at asymptotic early and late times, but also during the perturbation. Na\"ively, this is not a well-defined notion for such a non-equilibrium process, as the particle number at intermediate times depends on a basis choice of reference states with respect to which particles and anti-particles are defined, even though the final late-time particle number is independent of this basis choice. The basis choice is associated with a particular truncation of the adiabatic expansion. The adiabatic expansion is divergent, 
and we show that if this divergent expansion is truncated at its optimal order, a universal time dependence is obtained, confirming a general result of Dingle and Berry. This optimally truncated particle number provides a clear picture of quantum interference effects for perturbations with non-trivial temporal sub-structure. We illustrate these results using several equivalent definitions of adiabatic particle number: the Bogoliubov, Riccati, Spectral Function and Schr\"odinger picture approaches. In each approach, the particle number may be expressed in terms of the tiny deviations between the exact and adiabatic solutions of the Ermakov-Milne equation for the associated time-dependent oscillators.

\end{abstract}



\maketitle

\section{Introduction}

The stimulated production of particles from the quantum vacuum is a remarkable feature of quantum field theory that can occur when the vacuum is subjected to an external perturbation, such as gauge or gravitational curvature. Notable examples include the Schwinger effect from applying an external electric field to the quantum electrodynamic (QED) vacuum \cite{he,sch,greiner,dunne}, Friedmann-Robertson-Walker (FRW) cosmologies \cite{parker68,zeldovich,Parker:1974qw,mukhanov,hu}, de Sitter space times \cite{Birrell:1982ix,mott,bousso,gds2010,Polyakov:2007mm,and}, Hawking Radiation due to blackholes and gravitational horizon effects \cite{gibbhawk,vacha,ford,Boyanovsky:1996rw,Brout:1995rd,padma}, and Unruh Radiation seen by an accelerating observer \cite{unruh,Schutzhold:2006gj}. This particle production paradigm plays an important role in the physics of non-equilibrium processes in heavy-ion collisions \cite{gelisvenu,kharzeevtuchin,Gelis:2010nm}, astrophysical phenomena \cite{ruff}, and the search for nonlinear and non-perturbative effects in ultra-intense laser systems \cite{mourou,mattias,dunne-eli,DiPiazza:2011tq}. There are also close technical analogues with driven two-level systems, relevant for atomic and condensed matter processes \cite{oka,Nation}, such as Landau-Zener-St\"uckelberg transitions \cite{lzs}, the dynamical Casimir effect and its analogues \cite{Dodonov:2010zza,nori}, Ramsey processes and tunnel junctions \cite{AkkermansDunne:2012,reulet}. 

Particle production involves  evolution of a quantum system from an initial (free) equilibrium configuration to a new final (free) equilibrium configuration through an intervening non-equilibrium evolution due to a perturbing background. Quantifying the final asymptotic particle number involves relating the final equilibrium configuration to the initial one. This is a comparison of well-defined asymptotic vacua where the identification of positive (particles) and negative (anti-particles) energy states is unambiguous and exact. 
On the other hand, a quantitative description of particle production at all times, not just at asymptotically early and late times, requires a well-defined notion of time-dependent particle number also at {\it intermediate times}. This is a challenging conceptual and computational problem, especially if one wants to include also back-reaction effects and the full non-equilibrium dynamics. In this paper we discuss in detail one significant aspect of this problem: the role of the truncation of the adiabatic expansion in the conventional definition of time-dependent particle number. 

At intermediate times, when the system is out of equilibrium, it is less clear how to distinguish between positive and negative energy states. The standard approach \cite{parker68,mukhanov,hu,brezin,popov,gavrilov,KME:1998,Habib:1999cs,kimpage,winitzki,Kim:2011jw,Gelis:2015kya,Zahn:2015awa} involves using the adiabatic expansion to specify a reference basis set of approximate states, under the assumption of a slowly varying perturbation. Then a time-dependent particle number is defined by the projection of the evolving system onto these approximate states.  With this procedure, the final particle number at asymptotically late time is independent of the basis choice.  However, the particle number at intermediate times has a significant dependence on the basis choice, often varying over several orders of magnitude before settling down to its final basis-independent late-time value \cite{DabrowskiDunne:2014}. At first sight, this basis dependence would seem to immediately invalidate any attempt to define a physically sensible intermediate-time particle number. In particular, since the adiabatic expansion is a divergent expansion, we expect that its truncation should be performed at its optimal order, which is not fixed at a particular order but depends on the physical parameters of the perturbation. But here we can invoke a remarkable universality result due to Dingle  and Berry. Dingle found that the large-order behavior of the divergent adiabatic expansion has a universal form, providing accurate estimates of its behavior under optimal truncation \cite{dingle}. Berry \cite{BerryAsymptotic} applied Borel summation to find a generic smoothing of the associated Stokes phenomenon [i.e., particle production \cite{dumludunne}], leading to a universal time evolution. We have previously applied these technical results to the physical phenomena of particle production in time dependent electric fields and in de Sitter space time \cite{DabrowskiDunne:2014}. Here we present a systematic analysis of the influence of the choice of order of truncation of the adiabatic expansion, which corresponds directly to the non-uniqueness of specifying the approximate adiabatic reference states.

This surprising universality suggests a natural definition of time-dependent adiabatic particle number at all times, corresponding to an \emph{optimal adiabatic approximation} of the time evolution. This raises interesting questions regarding the physical nature of such a definition of particle number, some aspects of which have begun to be tested experimentally in analogous non-relativistic quantum systems \cite{lim-berry,expberry,demirice,BB,delCampo,jarz-shortcut,CDexp1,CDexp2}. We will address these questions in the quantum field theory context in future wrok.

In this paper, we examine the truncation of the  adiabatic expansion using several common (and equivalent) formulations of particle production: the Bogoliubov \cite{brezin,popov,KME:1998,and}, Riccati \cite{dumludunne}, Spectral Function \cite{FukushimaHataya:2014,Fukushima:2014} and 
Schr\"odinger \cite{vacha,padma} approaches. The analysis also extends straightforwardly to the quantum kinetic approach \cite{KME:1998,Habib:1999cs,Rau:1995ea,schmidt,Huet:2014mta}, and the Dirac-Heisenberg-Wigner approach with time-dependent background fields \cite{Hebenstreit:2010vz,Hebenstreit:2010cc}. For definiteness we study the Schwinger effect in scalar QED (sQED) with spatially homogeneous but time-dependent electric fields, but the basic results apply to a wide variety of quantum systems, as mentioned above. In Section \ref{s:APN} we review the relation between the Klein-Gordon equation and the Ermakov-Milne \cite{steen,erm,milne,pinney} equation, associated with the exact solution to the quantum harmonic oscillator with time-dependent frequency \cite{husimi,LewisInvariant1,DittrichReuter}.
The projection of the adiabatic states onto the exact  solution of the Ermakov-Milne equation leads to an analytic expression for the time-dependent adiabatic particle number, which clearly illustrates the basis dependence and simplifies its evaluation. 
The four approaches to time-dependent particle production yield precisely the same form, demonstrating that basis dependence is a universal feature of the adiabatic particle number at intermediate times.  In Section \ref{s:SAPN} we examine the influence of different truncations of the adiabatic expansion. This also yields a new perspective: the adiabatic approximation of time-dependent particle production is completely characterized by the exponentially small deviations from the exact Ermakov-Milne solution. Section \ref{s:Conc} is devoted to a brief discussion of the results.

\section{Adiabatic  Particle Number}
\label{s:APN}

\subsection{Field Mode Decomposition: Klein-Gordon and Ermakov-Milne Equations}
\label{ss:MEr}

We consider scalar QED for simplicity.\footnote{Apart from the opposite phase of interference effects, the physics is very similar to that of spinor QED, but it is notationally simpler.} For a charged (complex) scalar field $\Phi$ in a time-dependent and spatially homogeneous classical electric field, the scalar field can be separated into spatial Fourier modes, $\Phi_k(t) \sim f_k(t) e^{i \vec{k}\cdot \vec{x}}$, so that the Klein-Gordon equation, $\left( D_\mu^2 + m^2 \right) \! \Phi = 0$, reduces to decoupled linear time-dependent oscillator equations:
\begin{align}
\text{Klein-Gordon equation:}\qquad\qquad \ddot{f}_k(t) + \omega_k^2(t) f_k(t) = 0
\label{e:EOM}
\end{align}
Here the effective time-dependent frequency $\omega_k(t)$  is \cite{brezin,popov,KME:1998}
\begin{eqnarray}
\omega_k^2(t) \equiv m^2 + k_\perp^2 + \left( k_\parallel - A_\parallel(t) \right)^2
\label{e:omega}
\end{eqnarray}
where $k_\parallel$ and $k_\perp$ are the momenta of the produced particles along and transverse to the direction of the electric field, respectively. The magnitude of the electric field varies with time as $E(t)=-\dot{A}_\parallel(t)$. There is an analogous mode decomposition for particle production in cosmological and gravitational backgrounds  \cite{parker68,and,gds2010,vacha}.

We define quantized scalar field operators $\phi_k(t)$ and momenta $\pi_k(t)$ for each mode as 
\begin{eqnarray}
\phi_k(t) &=& f_k(t) \, a_k + f_{-k}^*(t)\,  b_{-k}^\dag 
\label{e:phik}\\
\pi_k^\dag(t) &=& \dot{f}_k(t) \, a_k + \dot{f}_{-k}^*(t)\,  b_{-k}^\dag 
\label{e:pik}
\end{eqnarray}
with (time independent) bosonic creation and annihilation operators to describe particles and anti-particles. Bosonic commutation relations impose the Wronskian condition on the mode functions $f_k(t)$:
\begin{align}
f_k(t) \dot{f}_k^*(t) - \dot{f}_k(t) f_k^*(t) = i
\label{e:Wronk}
\end{align}
Writing the complex mode function $f_k(t)$ in terms of its real amplitude $\xi_k(t)$ and phase $\lambda_k(t)$,
\begin{align}
f_k(t) \equiv \xi_k(t) \; e^{- i \lambda_k(t)}
\label{e:xik}
\end{align}
the Klein-Gordon equation (\ref{e:EOM}) reduces to the Ermakov-Milne \cite{steen,erm,milne,pinney} equation for the amplitude function $\xi_k$:
\begin{align}
\text{Ermakov-Milne equation:}\qquad\qquad \ddot{\xi}_k(t) + \omega_k^2(t) \xi_k(t) = \frac{1}{4 \xi_k^3(t)}
\label{e:Erm}
\end{align}
As usual, unitarity determines the time-dependent phase $\lambda_k(t)$ in terms of $\xi_k(t)$ as:
\begin{align}
\lambda_k(t) = \int^t \! \! \frac{dt^\prime}{2 \xi_k^2(t^\prime)} \;\; .
\label{e:lk}
\end{align}
Note that with the definition (\ref{e:xik}), the Ermakov-Milne equations (\ref{e:Erm}, \ref{e:lk}) are completely equivalent to the original Klein-Gordon equation (\ref{e:EOM}).
Another equivalent way to express the time-evolution is achieved by defining the square of the amplitude function, $G_k(t)\equiv \xi_k^2(t)$, which satisfies a nonlinear second-order equation, and its corresponding  linear third-order equation:
\begin{eqnarray}
\text{Gel'fand-Dikii equation}:\qquad\qquad 2G_k\, \ddot{G}_k - \dot{G}_k^2+4\,\omega_k^2(t)\, G_k^2&=&1 \quad \text{(nonlinear form)}\\
\dddot{G}_k+4\, \omega_k^2(t)\,\dot{G}_k+4\, \omega_k(t)\, \dot{\omega}_k(t)\, G_k &=&0
\quad \text{(linear form)}
\label{e:gd}
\end{eqnarray}
This is known as the Gel'fand-Dikii equation \cite{gelfand}, arising in the analysis of the resolvent Green's function for Schr\"odinger operators, which can be written in terms of products of solutions to the Klein-Gordon equation (\ref{e:EOM}).
 The resolvent approach has been used in the analysis of Schwinger effect \cite{Balantekin:1990aa,dunnehall}. 

\begin{figure}[!t]
\centering
\includegraphics[width=\textwidth]{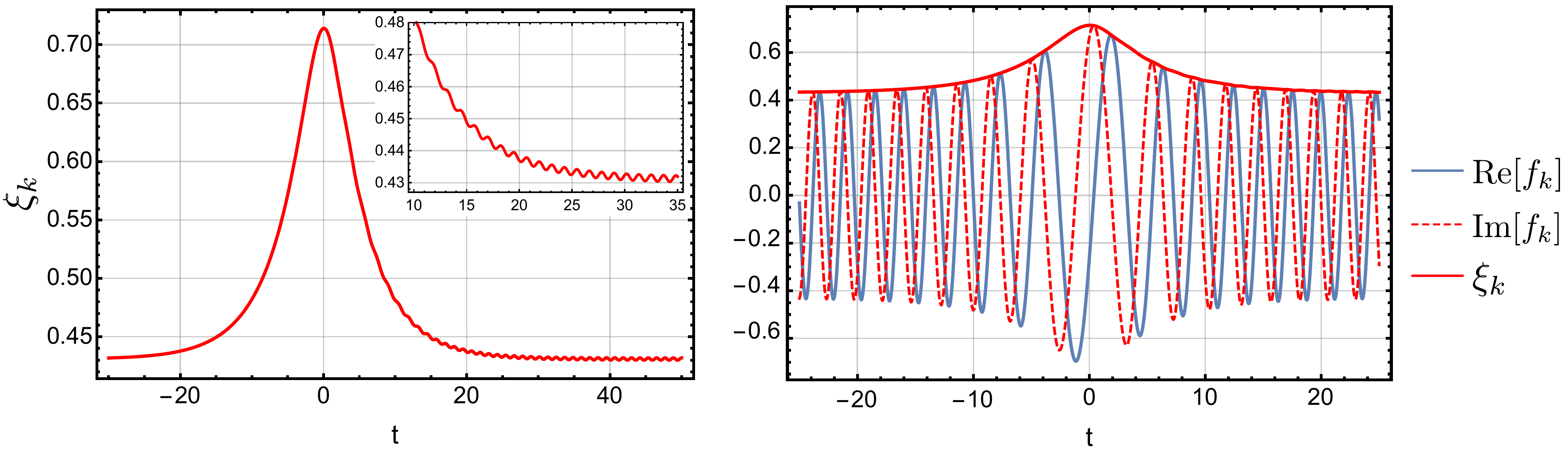}
\caption{ 
Plots of the amplitude function $\xi_k$ (left), 
and the real (blue-solid line) and imaginary (red-dashed line) parts of the mode function $f_k$ (right), 
with the scattering boundary conditions appropriate for the particle-production problem, 
for a time-dependent single-pulse electric field given by $E(t) = E_0 \text{sech}^2(a t)$, with  magnitude $E_0 = 0.25$, $a = 0.1$, longitudinal momentum $k_\parallel = 0$, and transverse momentum $k_\perp$ = 0, all in units with $m=1$.
For this electric field, both $f_k$ and $\xi_k$ can be obtained analytically (see Appendix \ref{s:Ex1P}), and $\xi_k$ is plotted as a solid-red line in each subplot for comparison.
Note the smooth behavior of $\xi_k(t)$, with small oscillations about the final asymptotic value $1/\sqrt{2\omega_k(+\infty)}$ shown in the inset figure on the left. As we show in this paper, these small oscillations encode the particle production phenomenon.}
\label{f:fxi}
\end{figure}

The particle production problem consists of the following physical situation: at initial time the vacuum is defined with respect to the (time-independent) creation and annihilation operators in (\ref{e:phik}). Then as time evolves the vacuum is subjected to a time-dependent electric field, which turns off again as $t\to+\infty$. At $t=+\infty$, after the electric field has been turned off, the production of particles from vacuum can be inferred from the fraction of negative frequency modes in the evolved mode functions. As is well known \cite{brezin,popov,dumludunne}, this can be expressed as an ``over-the-barrier''  quantum mechanical scattering problem, in the time domain, by interpreting the Klein-Gordon equation (\ref{e:EOM}) as a Schr\"odinger-like equation
\begin{eqnarray}
-\ddot{f}_k -\left(k_\parallel -A_\parallel(t)\right)^2 f_k = \left(m^2+k_\perp^2\right) f_k
\label{e:scattering}
\end{eqnarray}
with  physical ``scattering''  boundary conditions \cite{brezin,popov,gavrilov}:
\begin{align}
f_k(t) \sim 
\begin{dcases*}
       \tfrac{1}{\sqrt{2 \omega_k(-\infty)}} e^{- i \omega_k(-\infty) t}  & , $\quad t \to - \infty $\\
        \tfrac{1}{\sqrt{2 \omega_k(+\infty)}} \left( A_k e^{- i \omega_k(+\infty) t} +  B_k e^{i \omega_k(+\infty) t} \right) & , $\quad t \to + \infty$
        \end{dcases*}
        \label{e:Scat}
\end{align}
The scattering coefficients $A_k$ and $B_k$ defined at $t=+\infty$ satisfy $|A_k|^2-|B_k|^2=1$. So, we can evolve
the mode oscillator equation (\ref{e:EOM}) with initial conditions
\begin{align}
f_k(t \to - \infty) &\sim \frac{1}{\sqrt{2 \omega_k(-\infty)}} e^{- i \omega_k(-\infty) t}		\\
\dot{f}_k(t \to - \infty) & \sim -i\sqrt{ \frac{\omega_k(-\infty)}{2}} e^{- i \omega_k(-\infty) t}	
\end{align}
or, equivalently the Ermakov-Milne equation (\ref{e:Erm}) with initial  conditions
\begin{align}
\xi_k(t \to - \infty) &\sim  \frac{1}{\sqrt{2 \omega_k(-\infty)}}		
\label{e:BCxi}		\\
\dot{\xi}_k(t \to - \infty) &\sim  0
\label{e:BCdxi}
\end{align}
A numerical advantage of the Ermakov-Milne equation is that the amplitude function $\xi_k(t)$ typically varies more smoothly than the mode function $f_k(t)$ [and recall from (\ref{e:lk}) that the phase $\lambda_k(t)$ is determined by $\xi_k(t)$]. This is illustrated in Figure \ref{f:fxi}, for an explicit example of a single-pulse electric field, for which a well-known analytic exact solution is possible, as reviewed in the Appendix \ref{s:Ex1P}. 
 In this paper we primarily express particle number in terms of the amplitude function $\xi_k(t)$. 

\subsection{Bogoliubov Transformation and Adiabatic Particle Number}
\label{ss:BTA}

In processes that involve a time-dependent background field,  a unique separation into positive and negative energy states with which to identify particles and anti-particles is only possible at asymptotic times \cite{brezin,popov}, when the electric field is turned off. This is the same as the non-uniqueness of defining left- and right-moving modes inside an inhomogeneous dielectric medium \cite{budden,berry-mount}.

To proceed, we define a time-dependent {\it adiabatic} particle number in the presence of a slowly varying time-dependent background, with respect to a particular set of reference mode functions $\tilde{f}_k(t)$ defined as 
\begin{align}
\tilde{f}_k(t)
\equiv \frac{1}{\sqrt{2 W_k}} e^{- i \int^t W_k(t)}
\xrightarrow{t \to -\infty} \frac{1}{\sqrt{2 \omega_k(-\infty)}} e^{- i \omega_k(-\infty) t } 
\label{e:refm}
\end{align}
Clearly there is an infinite number of such reference mode functions, all having the same initial asymptotic behavior. The problem is to define a physically suitable set of mode functions for use at intermediate times.

Insisting that $\tilde{f}_k$, as defined in (\ref{e:refm}), be a solution to the Klein-Gordon  equation (\ref{e:EOM}), the function $W_k(t)$ is related to the effective frequency $\omega_k(t)$ by the well-known Schwarzian derivative form:
\begin{align}
W_k^2(t) = \omega_k^2(t) - \left[ \frac{\ddot{W}_k(t)}{2 W_k(t)} - \frac{3}{4} \left( \frac{\dot{W}_k(t)}{W_k(t)} \right)^2 \right]
\label{e:Wcon}
\end{align}
This can be solved by a systematic adiabatic expansion in which the leading order is the standard leading WKB solution to the mode oscillator equation (\ref{e:EOM}) of the form $W_k^{(0)}(t) = \omega_k(t)$ \cite{BerryAsymptotic,DabrowskiDunne:2014}. Higher order terms are analyzed in detail in section \ref{s:SAPN}.

The Bogoliubov Transformation is a linear canonical transformation that defines a set of time-dependent creation and annihilation operators, $\tilde{a}_k(t)$ and $\tilde{b}_{-k}(t)$, from the original time-independent operators, $a_k$ and $b_{-k}$, defined at the initial time in (\ref{e:phik}, \ref{e:pik}) \cite{popov}. 
They are related by
\begin{align}
\begin{pmatrix}
\tilde{a}_k(t)		\\
\tilde{b}_{-k}^\dag(t)
\end{pmatrix}
=
\begin{pmatrix}
\alpha_k(t)	& \beta_k^*(t)		\\
\beta_k(t)		& \alpha_k^*(t)
\end{pmatrix}
\begin{pmatrix}
a_k			\\
b_{-k}^\dag
\end{pmatrix}
\label{e:Bog}
\end{align}
where unitarity requires $\left| \alpha_k(t) \right|^2 - \left| \beta_k(t) \right|^2 = 1$ for scalar fields, for all $t$.
As a result of the Bogoliubov transformation, the equivalent decomposition of the scalar field operator in terms of these reference mode functions is
\begin{align}
\phi_k(t) =  \tilde{f}_k(t)\, \tilde{a}_k(t) + \tilde{f}_k^*(t)\,  \tilde{b}_{-k}^\dag(t)
\label{e:bogphi}
\end{align}
This can also be interpreted as a linear transformation between the exact mode functions $f_k(t)$ and the reference adiabatic mode functions $\tilde{f}_k(t)$, as
\begin{align}
f_k(t) = \alpha_k(t) \tilde{f}_k(t) + \beta_k(t) \tilde{f}_k^*(t)
\label{e:fk}
\end{align}
We also need to specify the transformation of  the scalar field momentum operator $\pi_k^\dag = \dot{\phi}_k$:
\begin{align}
\pi_k^\dag(t) = Q_k(t) \tilde{f}_k(t)\,  \tilde{a}_k(t) + Q_k^*(t)  \tilde{f}_k^*(t)\, \tilde{b}_{-k}^\dag(t)
\label{e:bogpi}
\end{align}
with a corresponding decomposition of the first derivative:
\begin{align}
\dot{f}_k(t) = Q_k(t) \alpha_k(t) \tilde{f}_k(t) + Q_k^*(t) \beta_k(t) \tilde{f}_k^*(t)
\label{e:dfk}
\end{align}
Here $Q_k(t)$ is  defined as 
\begin{align}
Q_k(t) = - i W_k(t) + V_k(t) \;\; .
\label{e:Qk}
\end{align}
The inclusion of the real time-dependent function $V_k(t)$, specified later, in the decompositions (\ref{e:fk}) and (\ref{e:dfk}) represents the most general decomposition of the exact solution $f_k$ that is consistent with unitarity (the preservation of the bosonic commutation relations, or equivalently the Wronskian condition (\ref{e:Wronk})). The freedom in the choice of $W_k(t)$ and $V_k(t)$ encodes the arbitrariness of specifying positive and negative energy states at intermediate times. We will see later that a `natural' choice is $V_k=-\dot{W}_k/(2W_k)$, coming from the derivative of the $1/\sqrt{2W_k}$ factor in the definition of the reference mode functions (\ref{e:refm}).

The scattering coefficients in (\ref{e:Scat}) are realized as the Bogoliubov coefficients evaluated at asymptotically late time, after the perturbation has turned off: $A_k = \alpha_k(+\infty)$ and $B_k = \beta_k(+\infty)$. The time-dependent adiabatic particle number, for each mode $k$, is defined as the expectation value of the time-dependent number operator $\tilde{a}_k^\dag(t) \tilde{a}_k(t)$  with respect to the asymptotic vacuum state. 
Assuming no particles are initially present, the time-dependent adiabatic particle number is
\begin{align}
\tilde{\mathcal{N}}_k(t) \equiv \left\langle \tilde{a}_k^\dag(t) \tilde{a}_k(t) \right\rangle = \left| \beta_k(t) \right|^2 \;\; .
\label{e:APN}
\end{align}
This reduces the problem to the direct evaluation of the time evolution of the Bogoliubov transformation parameters $\alpha_k(t)$ and $\beta_k(t)$. 
The decompositions (\ref{e:fk}) and (\ref{e:dfk}) are exact provided they satisfy the mode oscillator equation (\ref{e:EOM}), which implies the following evolution equations for the Bogoliubov transformation parameters $\alpha_k(t)$ and $\beta_k(t)$:
\begin{align}
\begin{pmatrix}
\dot{\alpha}_k		\\
\dot{\beta}_k
\end{pmatrix}
=
\begin{pmatrix}
\delta_k	& \left( \Delta_k + \delta_k \right) e^{2 i \int^t W_k}	\\
\left( \Delta_k + \delta_k^* \right) e^{- 2 i \int^t W_k}		& \delta_k^*
\end{pmatrix}
\begin{pmatrix}
\alpha_k		\\
\beta_k
\end{pmatrix}
\label{e:ABevo}
\end{align}
where
\begin{align}
\delta_k &= \frac{1}{2 i W_k} \left( \omega_k^2 - W_k^2 + \left( \dot{V}_k + V_k^2 \right) \right) 
\label{e:dk} \\
\Delta_k &= \frac{\dot{W}_k}{2 W_k} + V_k
\label{e:Dk}
\end{align}
Note that $\delta_k$ vanishes with the choice $V_k=-\dot{W}_k/(2 W_k)$. 
The numerical evaluation of this coupled differential equation completely determines the time evolution of $\alpha_k(t)$ and $\beta_k(t)$ with respect to the basis $(W_k,V_k)$.
The time evolution of the adiabatic particle number $\tilde{\mathcal{N}}_k(t)$ is obtained by the modulus squared of the time evolution of the Bogoliubov coefficient following (\ref{e:APN}),  solved using the initial conditions $\alpha_k(-\infty) = 1$ and $\beta_k(-\infty) = 0$, consistent with the scattering scenario in (\ref{e:Scat}) and the assumption of no particles being initially present.
The evolution equations (\ref{e:ABevo}) are dependent on the choice made for the basis functions $W_k(t)$ and $V_k(t)$, which influences the time evolution of the adiabatic particle number at intermediate times but does not affect its final asymptotic value at future infinity, $\left| B_k \right|^2$ \cite{DabrowskiDunne:2014}. This is because the final value is determined by the {\it global} information of the Stokes phenomenon \cite{dumludunne}.

The time evolution of the coefficients $\alpha_k(t)$ and $\beta_k(t)$ can also be expressed directly through the time evolution of the amplitude function $\xi_k(t)$.
Solving the linear equations (\ref{e:fk}) and (\ref{e:dfk}) we find 
\begin{align}
\alpha_k(t) &= i \tilde{f}_k^*(t) \left( \dot{f}_k(t) - Q_k^*(t) f_k(t) \right) 	
\label{e:Na}	\\
\beta_k(t) &= - i \tilde{f}_k(t) \left( \dot{f}_k(t) - Q_k(t) f_k(t) \right) \quad ,
\label{e:Nb}
\end{align}
Furthermore, from (\ref{e:xik}) and its time-dependent phase (\ref{e:lk}), we find the identity
\begin{align}
\frac{\dot{f}_k}{f_k} = \frac{\dot{\xi}_k}{\xi_k} - \frac{i}{2 \xi_k^2} \quad ,
\label{e:INfxi}
\end{align}
Thus, the Bogoliubov coefficients may be rewritten in the uncoupled form as
\begin{align}
\alpha_k(t) &= \frac{\xi_k}{\sqrt{2 W_k}} 
\left[ \left(\frac{1}{2 \xi_k^2} + W_k \right) + i \left( \frac{\dot{\xi}_k}{\xi_k} - V_k \right)  \right]  
\exp\left[- i \int^t \left( \frac{1}{2 \xi_k^2} - W_k \right)\right] 		\\
\beta_k(t) &= - \frac{\xi_k}{\sqrt{2 W_k}} 
\left[ \left( \frac{1}{2 \xi_k^2} - W_k \right) + i \left( \frac{\dot{\xi}_k}{\xi_k}  - V_k \right)  \right] 
\exp\left[- i \int^t \left( \frac{1}{2 \xi_k^2} + W_k \right)\right] 
\label{e:Bk}
\end{align}
This expresses the time evolution of the Bogoliubov coefficients as a comparison between the time evolution of the amplitude function, $\xi_k(t)$, obtained by solving the Ermakov-Milne equation (\ref{e:Erm}), and the reference mode basis $(W_k,V_k)$.
The Adiabatic Particle Number then follows:
\begin{align}
\left| \alpha_k(t) \right|^2 &= \frac{\xi_k^2}{2 W_k} \left[ \left( \frac{1}{2 \xi_k^2} + W_k \right)^2 + \left( \frac{\dot{\xi}_k}{\xi_k} - V_k \right)^2 \right] 
	\\
\tilde{\mathcal{N}}_k(t) = \left| \beta_k(t) \right|^2 &= \frac{\xi_k^2}{2 W_k} \left[ \left( \frac{1}{2 \xi_k^2} - W_k \right)^2 + \left( \frac{\dot{\xi}_k}{\xi_k} - V_k \right)^2 \right]
\label{e:APNB}
\end{align}
It is straightforward to confirm that unitarity is preserved: $\left| \alpha_k(t) \right|^2 - \left| \beta_k(t) \right|^2 = 1$. 

The expression (\ref{e:APNB}) for the time-dependent particle number is one of the primary results of this paper. It emphasizes clearly the dependence of the adiabatic particle number on the basis choice of reference mode functions $(W_k,V_k)$. It is not enough to know the time evolution of $\xi_k(t)$: one must also compare it to the reference functions. With the choice $V_k=-\dot{W}_k/(2W_k)$, the expression for the adiabatic particle number simplifies further to a direct comparison between $\xi_k(t)$ and $W_k(t)$:
\begin{align}
\tilde{\mathcal{N}}_k(t) &= \frac{\xi_k^2}{2 W_k} \left[ \left( \frac{1}{2 \xi_k^2} - W_k \right)^2 + \left( \frac{\dot{\xi}_k}{\xi_k} +\frac{\dot{W}_k}{2 W_k}\right)^2 \right]
\label{e:APNB2}
\end{align}

 In subsequent sub-sections we show how exactly the same expression arises in other  different but equivalent, methods for defining and computing the adiabatic particle number. Then in Section \ref{s:SAPN} we show how in the adiabatic expansion the expression (\ref{e:APNB2}) can be viewed as a measure of the tiny deviations between the exact solution $\xi_k(t)$ of the Ermakov-Milne equation and various orders of the adiabatic approximation for $W_k(t)$.

\subsection{Riccati Approach to Adiabatic Particle Number}
\label{ss:Rk}

The time evolution of the Bogoliubov coefficients can be re-expressed in Riccati form by defining the ratio \cite{popov,dumludunne}
\begin{align}
R_k(t) \equiv \frac{\beta_k(t)}{\alpha_k(t)}
\label{e:rk}
\end{align}
which can be viewed as a local (in time) reflection amplitude for this Schr\"odinger-like equation (\ref{e:scattering}) \cite{popov,brezin}. Using the unitarity condition, $\left| \alpha_k(t) \right|^2 - \left| \beta_k(t) \right|^2 = 1$, the time-dependent adiabatic particle can be rewritten as
\begin{align}
\tilde{\mathcal{N}}_k(t) = \frac{\left| R_k(t) \right|^2}{1 - \left| R_k(t) \right|^2}
\label{e:NR2}
\end{align}
In the semi-classical limit in which $m$ is the dominant scale (as is relevant in QED), this over-the-barrier scattering problem has an exponentially small reflection probability, which implies that
the adiabatic particle number is well approximated by $\tilde{\mathcal{N}}_k(t) \simeq \left| R_k(t) \right|^2$.

Using (\ref{e:rk}), the Bogoliubov coefficient evolution equations (\ref{e:ABevo}), with the basis ($W_k,V_k)$,  become a Riccati equation:
\begin{align}
\dot{R}_k = \left( \Delta_k - \delta_k \right) e^{- 2 i \int^t W_k} - 2 \delta_k R_k - \left( \Delta_k + \delta_k \right) e^{2 i \int^t W_k} R_k^2
\label{e:Rkeq}
\end{align}
with $\delta_k(t)$ and $\Delta_k(t)$ defined by equations (\ref{e:dk}, \ref{e:Dk}). This is straightforward to  evaluate numerically with the initial conditions $R_k(-\infty) = 0$, and an initial phase of zero. It can also be solved semiclassically for $R_k(+\infty)$, thereby yielding the final particle number $\tilde{\mathcal N}_k(+\infty)$, using complex turning points and the Stokes phenomenon \cite{dumludunne}.

Alternatively, using  the forms calculated previously for $\alpha_k(t)$ and $\beta_k(t)$, equations (\ref{e:Bk}), we obtain an analytic representation of  the reflection probability as
\begin{align}
\left| R_k \right|^2 = \frac{\left( \frac{1}{2 \xi_k^2} - W_k \right)^2 + \left( \frac{\dot{\xi}_k}{\xi_k} - V_k \right)^2}{\left( \frac{1}{2 \xi_k^2} + W_k\right)^2 + \left( \frac{\dot{\xi}_k}{\xi_k} - V_k \right)^2}
\label{e:RK}
\end{align}
Expression (\ref{e:NR2})  for the adiabatic particle number then yields
\begin{align}
\tilde{\mathcal{N}}_k(t) = \frac{\xi_k^2}{2 W_k} \left[ \left( \frac{1}{2 \xi_k^2} - W_k \right)^2 + \left( \frac{\dot{\xi}_k}{\xi_k} - V_k \right)^2 \right]
\label{e:APNR}
\end{align}
confirming the consistency with the Bogoliubov transformation expression (\ref{e:APNB}).

\subsection{Spectral Function Approach to Adiabatic Particle Number}
\label{ss:Spec}

Another physically interesting formalism to describe particle production at intermediate times is to define the time-dependent adiabatic particle number through the use of Spectral Functions \cite{Fukushima:2014,FukushimaHataya:2014}, which are constructed in terms of correlation functions of the time-dependent creation and annihilation operators (\ref{e:Bog}) used in (\ref{e:APN}). In this Section we show how the basis dependence arises in this formalism.

The Spectral Approach defines the adiabatic particle number through unequal time correlators of time-dependent creation and annihilation operators, in a limit that recovers the equal-time adiabatic particle number:
\begin{align}
\tilde{\mathcal{N}}_k(t) = \lim_{t_1,t_2 \to t} \left\langle \tilde{a}_k^\dag(t_1) \tilde{a}_k(t_2) \right\rangle
\label{e:APN3}
\end{align}
Using (\ref{e:Bog}, \ref{e:bogphi}), the time-dependent creation and annihilation operators can be written in terms of the decomposed field operators as
\begin{align}
\tilde{a}_{k}(t) &= i \tilde{f}_k^*(t) \left[ \partial_0 - Q_k^*(t) \right] \phi_k(t)			\\
\tilde{b}_{-k}^\dag(t) &= - i \tilde{f}_k(t) \left[ \partial_0 - Q_k(t) \right] \phi_k(t)
\end{align}
which match smoothly to the initial creation and annihilation operators. Note the dependence on the choice of basis $(W_k(t),V_k(t))$, through the function $Q_k(t)\equiv -i W_k(t)+V_k(t)$, defined in (\ref{e:Qk}). 
We thus obtain
\begin{align}
\tilde{\mathcal{N}}_k(t) = \frac{1}{2 W_k(t)} \lim_{t_1,t_2 \to t} \left( \left[ \partial_1 - Q_k(t_1) \right] \left[ \partial_2 - Q_k^*(t_2) \right] \right) \left\langle \phi_k^\dag(t_1) \phi_k(t_2) \right\rangle 	\quad ,
\label{e:APN3a}
\end{align}
where $\partial_j$ denotes a derivative with respect to time $t_j$. This expression shows a clear separation between the computation of the correlation function $\left\langle \phi_k^\dag(t_1) \phi_k(t_2) \right\rangle$, and the projection onto a  set of reference modes, characterized by $Q_k(t)$ in (\ref{e:Qk}). In \cite{Fukushima:2014,FukushimaHataya:2014} a particular basis choice was made,  $W_k=\omega_k$ and $V_k=0$, corresponding to a leading-order adiabatic expansion and a particular phase choice via $V_k$.  (\ref{e:APN3a}) makes it clear that this is just one of an infinite set of possible choices, for which the final particle number at late asymptotic time is always the same, but for which the particle number at intermediate times can be very different.

Spatially homogeneous time-dependent external electric fields decouple the modes $k$ allowing the spectral functions, the Wigner transformed Pauli-Jordan function $\mathcal{A}_k(t,k_0)$ and Hadamard function $\mathcal{D}_k(t,k_0)$, to be expressed as \cite{Fukushima:2014,FukushimaHataya:2014}
\begin{align}
\mathcal{A}_k(t,k_0) &= \frac{1}{\mathcal{V}} \int dT \, e^{i k_0 T} \left\langle \left[ \phi_k\!\left(t+\tfrac{T}{2}\right) , \phi_k^\dag\!\left( t - \tfrac{T}{2} \right)  \right] \right\rangle 
\label{e:WA}	\\
\mathcal{D}_k(t,k_0) &= \frac{1}{\mathcal{V}} \int dT \, e^{i k_0 T} \left\langle \left\{ \phi_k\!\left(t + \tfrac{T}{2} \right) , \phi_k^\dag\!\left( t - \tfrac{T}{2} \right)  \right\} \right\rangle
\label{e:WB}
\end{align}
with the conjugate variable pair being the energy $k_0$ and the time separation $T$. The spatial volume is denoted by $\mathcal{V}$.

The correlation function in (\ref{e:APN3a}) can be expressed through a linear combination of the inverse Wigner transformed functions (\ref{e:WA}, \ref{e:WB}) as
\begin{eqnarray}
\left\langle \phi_k^\dag\!\left(t - \tfrac{T}{2} \right) \! \phi_k\!\left(t + \tfrac{T}{2} \right) \right\rangle &=& \frac{\mathcal{V}}{2} \!\!
\int \frac{dk_0}{2 \pi} e^{- i k_0 T} \mathcal{W}_k(t,k_0) \\
&=&
\frac{\mathcal{V}}{2} \!\!
\int \frac{dk_0}{2 \pi} e^{- i k_0 T} \left(\mathcal{D}_k(t,k_0) - \mathcal{A}_k(t,k_0)\right)
\label{e:WW}
\end{eqnarray}
where the total spectral function is defined as $\mathcal{W}_k(t,k_0) \equiv \mathcal{D}_k(t,k_0) - \mathcal{A}_k(t,k_0)$.
Inserting this expression into (\ref{e:APN3a}), and taking the limit,  yields an expression for the time-dependent adiabatic particle number in terms of the transformed correlation function as
\begin{align}
\tilde{\mathcal{N}}_k(t) = \! \frac{\mathcal{V}}{4 W_k} \! \int \frac{dk_0}{2\pi} \bigg[ \frac{1}{4} \partial_t^2 - V_k \partial_t + \! \left( W_k + k_0 \right)^2 \!+ \!V_k^2 \bigg] 
\mathcal{W}_k(t,k_0)
\label{e:APN3b}
\end{align}
This expression (\ref{e:APN3b}) is the natural extension of Fukushima's result  \cite{Fukushima:2014,FukushimaHataya:2014}, which employed the leading adiabatic approximation choice of basis functions as $W_k(t) = \omega_k(t)$ and $V_k(t) = 0$, to a general basis specified by $W_k(t)$ and $V_k(t)$.
\begin{figure}[!ht]
\centering
\begin{align}
& \includegraphics[scale=0.45]{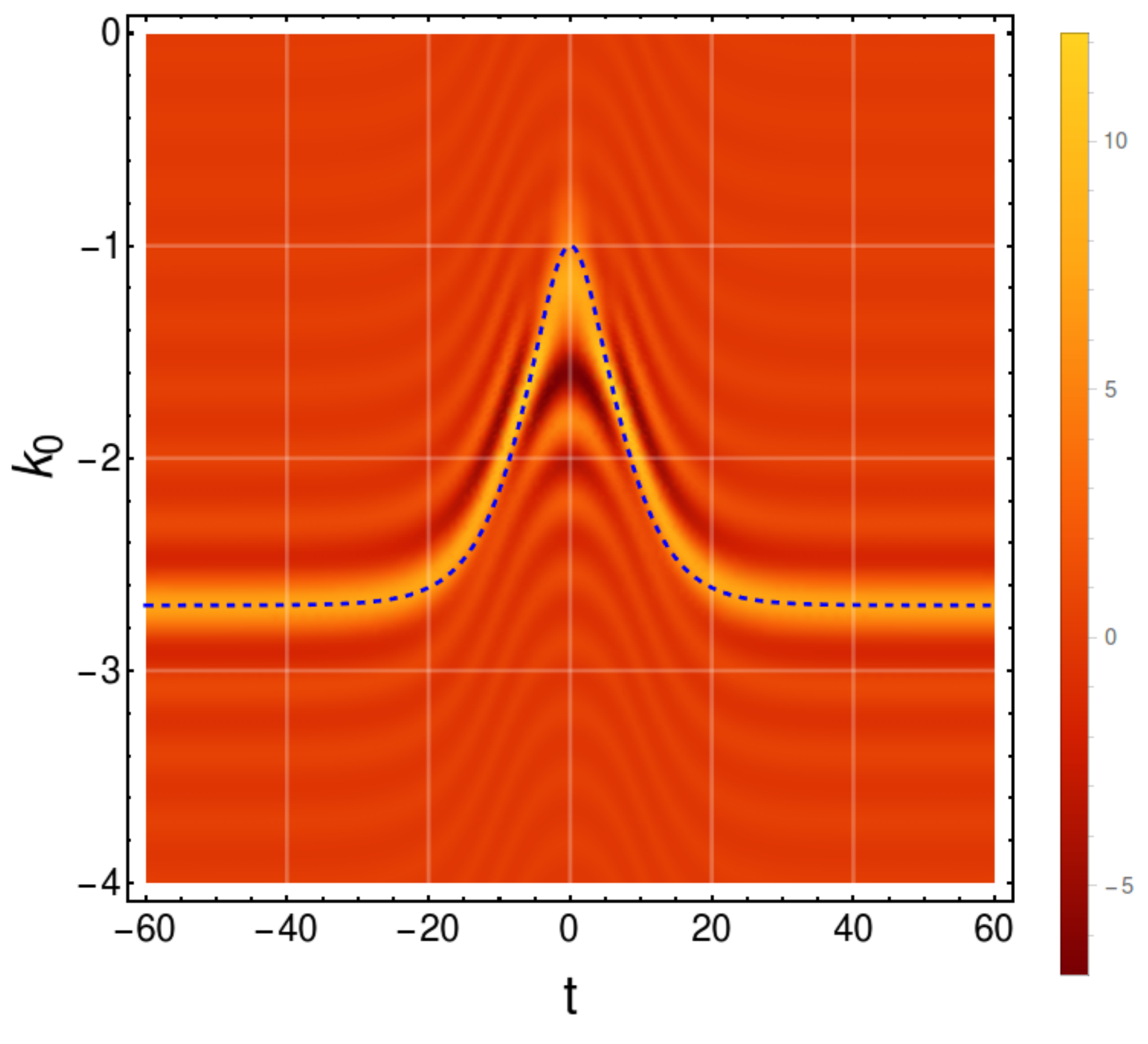} 
& \includegraphics[scale=0.45]{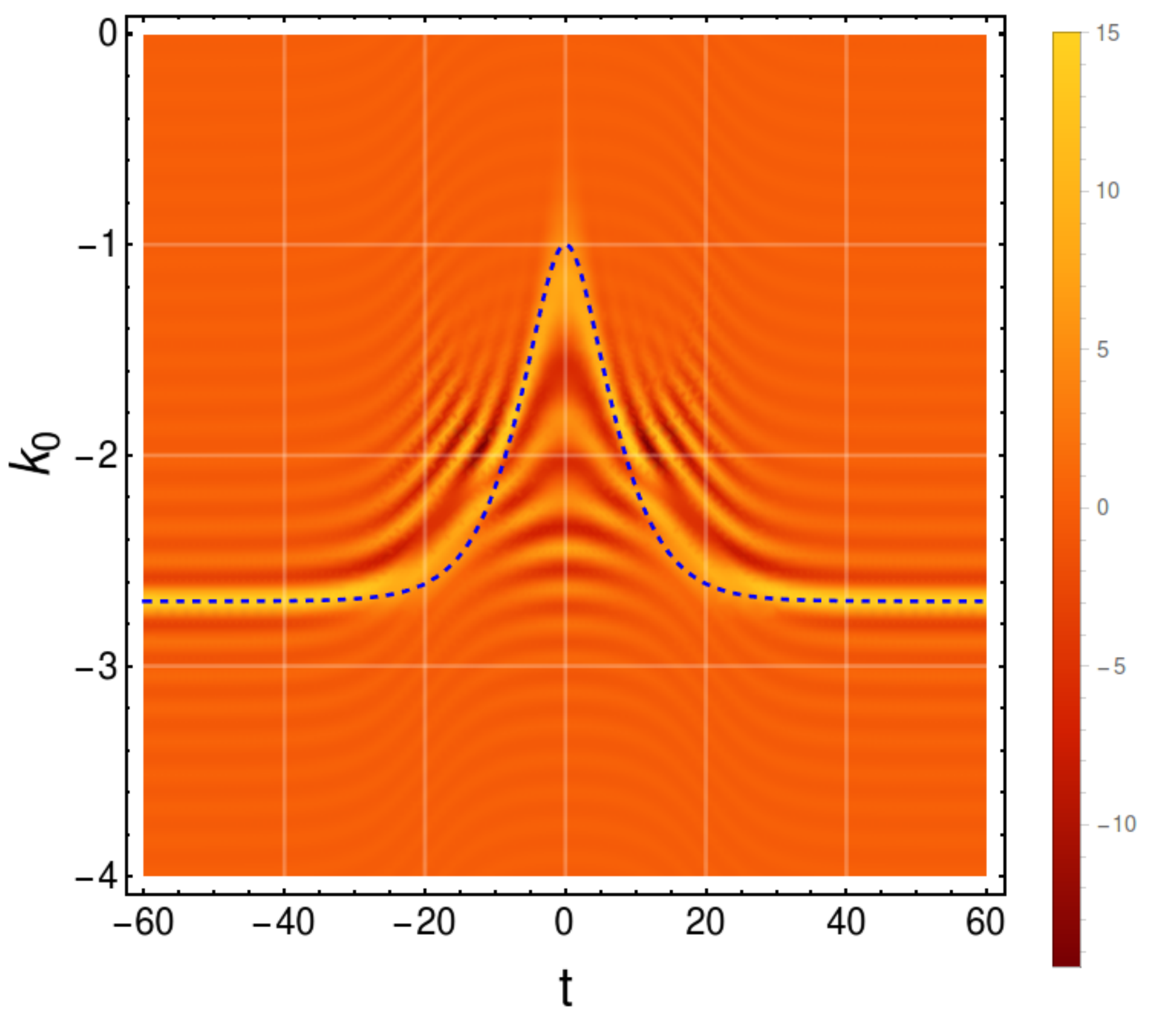} 
\nonumber \\
& \includegraphics[scale=0.45]{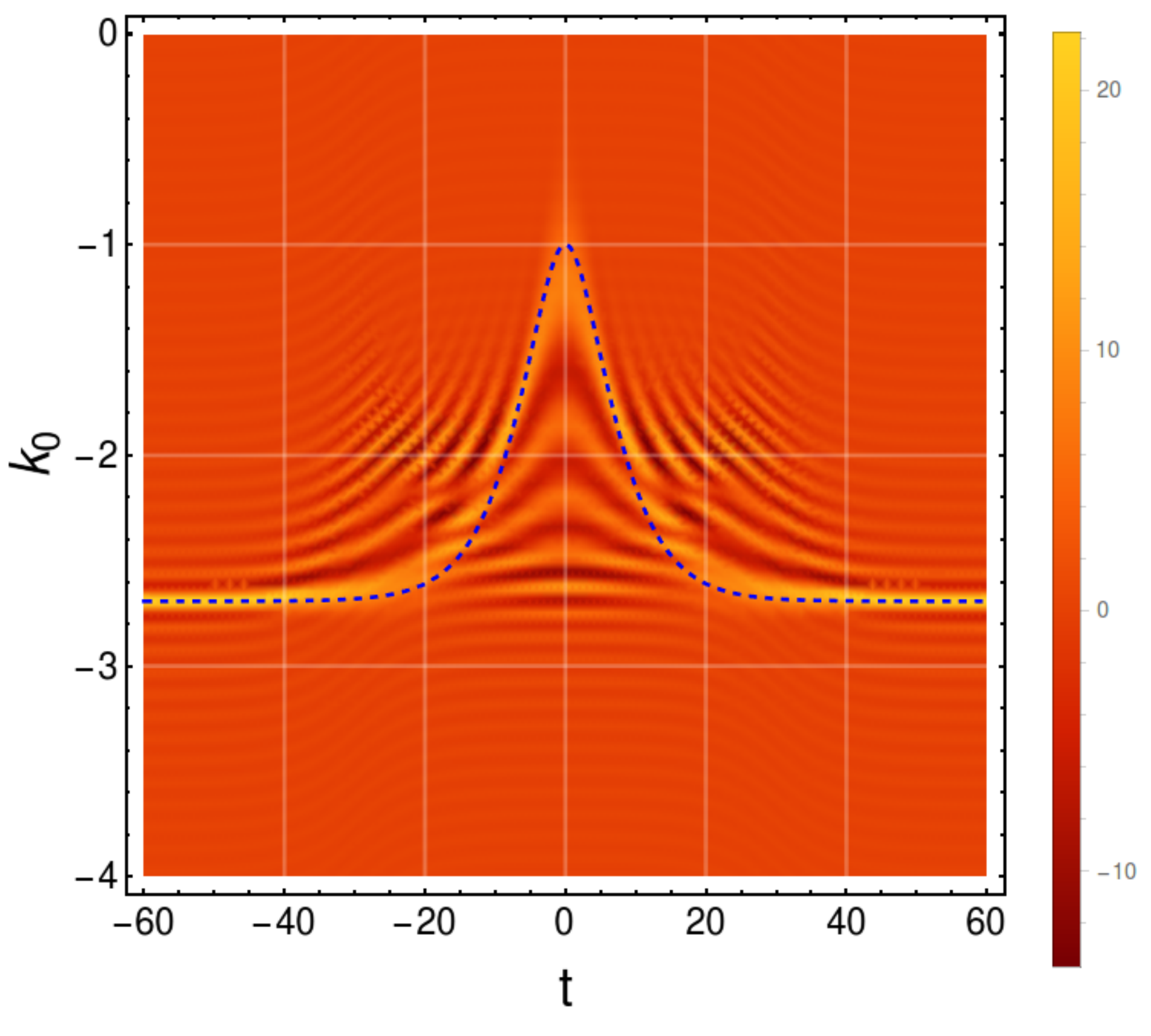} 
& \includegraphics[scale=0.45]{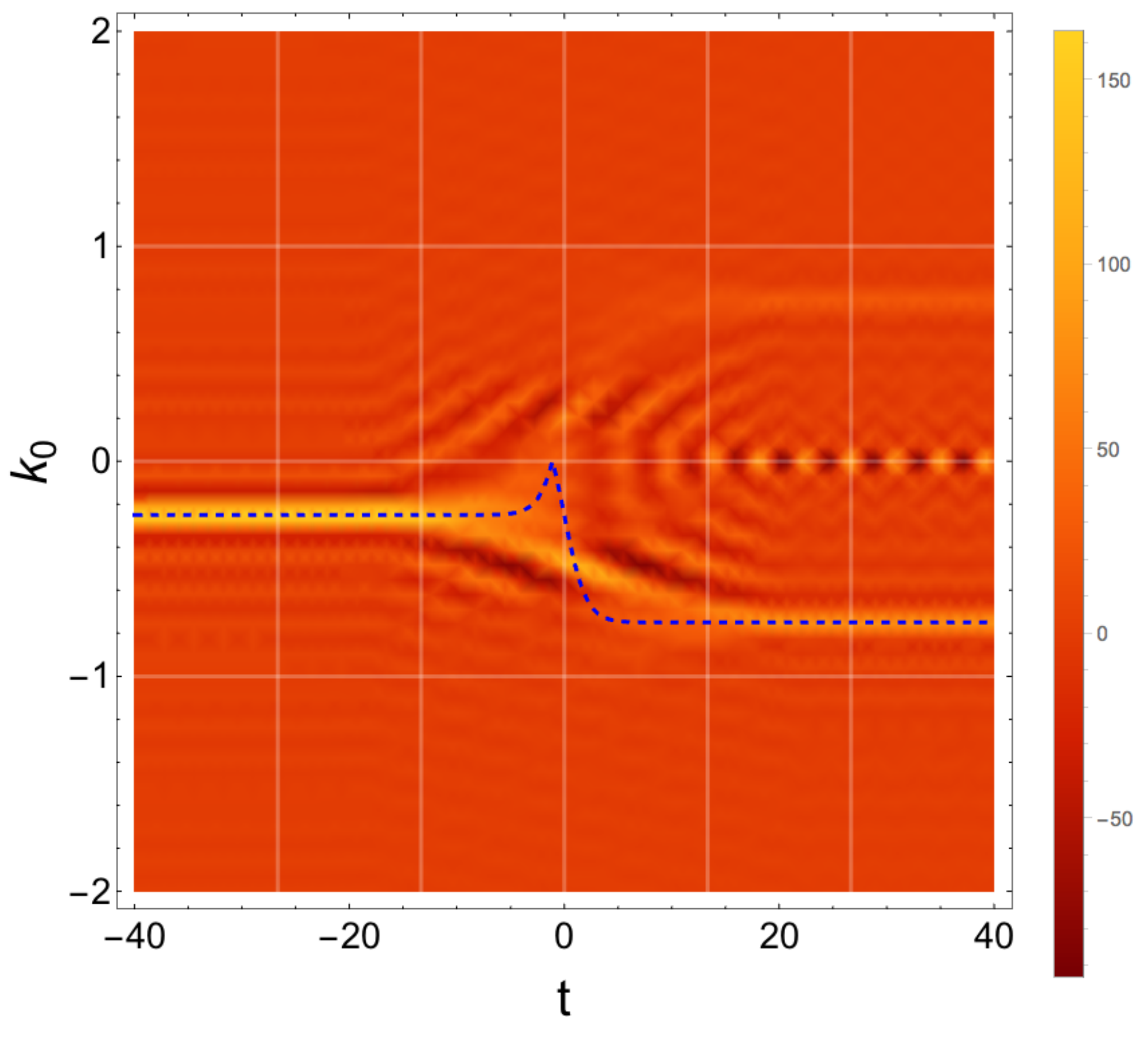} 
\nonumber
\end{align}
\caption{
Density plots with respect to $t$, and the conjugate energy variable $k_0$, of the spectral function $\mathcal{W}_k(t,k_0)$  for a time-dependent single-pulse electric field given by $E(t) = E_0 \text{sech}^2(at)$, obtained by numerically evaluating equation (\ref{e:WW3}) over the range $T=-T_0$ to $T=+T_0$, utilizing the exact solution $\xi_k(t)$ to the mode-oscillator equation found in the Appendix \ref{s:Ex1P}.  The upper left, upper right and lower left subplots are plotted with the magnitude $E_0 = 0.25, a = 0.1$, longitudinal momentum $k_\parallel = 0.25$, and transverse momentum $k_\perp = 0$, in units with $m=1$, with the upper left plot integrated with $T_0=20$, the upper right plot integrated with $T_0=40$, and the lower left subplot integrated with $T_0=60$. The lower right subplot is plotted for the physically unrealistic case with $m =0, E_0 = 0.5, k_\parallel = 0.25, k_\perp = 0$, as discussed in \cite{Fukushima:2014}, and integrated with $T_0=40$. In each subplot the dominant features of $\mathcal{W}_k(t,k_0)$ are well matched by the negative effective frequency, $-\omega_k(t)$ (\ref{e:omega}) (blue-dashed line), artificially plotted over each density subplot for direct comparison.
}
\label{f:spec}
\end{figure}

It is important to appreciate that the spectral function $\mathcal{W}_k(t,k_0)$ in (\ref{e:APN3b}) can be expressed directly in terms of the solutions to the Klein-Gordon equation or the Ermakov-Milne equation, without reference to the reference mode basis functions. Assuming no particles are initially present in the vacuum, the expectation value of the field operator commutator and anti-commutator are
\begin{align}
\left\langle \!\left[ \phi_k\!\left(t + \tfrac{T}{2} \right) , \phi_k^\dag\!\left( t - \tfrac{T}{2} \right)  \right] \!\right\rangle &= 
f_k\!\left(t + \tfrac{T}{2} \right) f_k^*\!\left(t - \tfrac{T}{2} \right) - f_k^*\!\left(t + \tfrac{T}{2} \right) f_k\!\left(t - \tfrac{T}{2} \right)
\\
\left\langle \!\left\{ \phi_k\!\left(t+\tfrac{T}{2} \right) , \phi_k^\dag\!\left( t - \tfrac{T}{2} \right)  \right\} \! \right\rangle &= 
f_k\!\left(t + \tfrac{T}{2} \right) f_k^*\!\left(t - \tfrac{T}{2} \right) + f_k^*\!\left(t + \tfrac{T}{2} \right) f_k\!\left(t - \tfrac{T}{2} \right)
\end{align}
Therefore, the spectral function $\mathcal{W}_k(t,k_0)$  assumes the form
\begin{align}
\mathcal{W}_k(t,k_0) = \frac{2}{\mathcal{V}} \! \int \! dT \, e^{i k_0 T} f_k\! \left( t - \tfrac{T}{2} \right) f_k^*\!\left(t + \tfrac{T}{2} \right)
\label{e:WW2}
\end{align}
Alternatively, this can be rewritten in terms of the amplitude function  $\xi_k(t)$:
\begin{align}
\mathcal{W}_k(t,k_0) = \frac{2}{\mathcal{V}} \! \int \! dT \, e^{i k_0 T} \xi_k\! \left( t - \tfrac{T}{2} \right) \xi_k\!\left(t + \tfrac{T}{2} \right) 
\text{exp}\!\left[ i \int^{t+T/2}_{t-T/2} \frac{dt^\prime}{2 \xi_k^2(t^\prime)} \right]
\label{e:WW3}
\end{align}
Thus, the spectral function $\mathcal{W}_k(t,k_0)$ is determined without any knowledge of the basis functions $(W_k(t),V_k(t))$ and is exact provided that integration is performed over all possible values of the separation $T$. The behavior of the spectral function (\ref{e:WW3}) is shown in Figure \ref{f:spec} for the soluble case of a single-pulse electric field (see Appendix \ref{s:Ex1P}), integrated over a finite range $T=-T_0$ to $T=+T_0$, for various values of the cutoff $T_0$.
The two upper subplots and the lower left subplots in Figure \ref{f:spec} are plotted for the case when $E_0=0.25, a=0.1, k_\perp= k_\parallel = 0$, in units with $m=1$, with the upper left plot integrated with $T_0=20$, the upper right plot integrated with $T_0=40$, and the lower left plot integrated with $T_0=60$. 
The lower right plot was plotted with the parameters  used in \cite{Fukushima:2014}, with $m=0, E_0=0.5, k_\parallel = 0.25, k_\perp = 0$ and integration with $T_0=40$. 
In each subplot of Figure \ref{f:spec}, the dominant features of $\mathcal{W}_k(t,k_0)$ (\ref{e:WW3}) are well approximated by the negative effective frequency $- \omega_k(t)$, plotted with a blue-dashed line, which demonstrates that the spectral function $\mathcal{W}_k(t,k_0)$ represents the projection of the fundamental frequency on a plane spanned by time and the conjugate energy variable $k_0$. 
Furthermore, we see that the oscillating features of the spectral function decrease as $T_0 \to \infty$.
Lastly, we compared the results obtained in \cite{Fukushima:2014}, calculated by numerically evaluating the mode function $f_k(t)$ and the subsequent integral in (\ref{e:WW3}), with the exact solution to the mode-oscillator equation (see Appendix \ref{s:Ex1P}), which indicates that the numerical approach suffers from sensitive numerical instabilities in the evaluation of (\ref{e:WW3}) and the mode function $f_k(t)$.

We next show how the expression for the time-dependent adiabatic particle number that was previously derived in the Bogoliubov (\ref{e:APNB}) and Riccati formalisms (\ref{e:APNR}) is obtained in the Spectral Representation formalism. From equation (\ref{e:APN3b}), and using the spectral function (\ref{e:WW2}), the expression is recovered by first re-writing the derivatives in terms of $t$, and reorganizing the resulting terms via integration by parts to eliminate, apart from the exponential term $e^{i k_0 T}$, the $k_0$ dependence in the integrand. The  $k_0$ integration produces a Dirac Delta function which, when integrated over $T$, eliminates all integrations. Two terms appear: one corresponding directly to the adiabatic particle number, and the other to a surface boundary term. Recast in terms of $\xi_k$ using the identity (\ref{e:INfxi}), this lengthy but straightforward calculation leads to an expression for the time-dependent adiabatic particle number (\ref{e:APN3b}) as
\begin{align}
\tilde{\mathcal{N}}_k(t) = \frac{\xi_k^2}{2 W_k} \left[ \left( \frac{1}{2 \xi_k^2} - W_k \right)^2 + \left( \frac{\dot{\xi}_k}{\xi_k} - V_k \right)^2 \right] 
\label{e:APNS}
\end{align}
noting that
the total surface boundary term vanishes when $T_0 \to \infty$.
This agrees precisely with the Bogoliubov and Riccati expressions in (\ref{e:APNB}). We see again that the adiabatic particle number is basis dependent at intermediate times, through the choice of the 
$W_k$ and $V_k$ functions.
As before, $\xi_k$ is solved exactly without any knowledge of the basis functions, and the selected basis functions are inserted into the expression (\ref{e:APN3a}) to determine the adiabatic particle number with respect to that basis. In the spectral function approach this follows because the spectral function (\ref{e:WW3})  is determined once and for all by the solution $\xi_k(t)$, and then the basis-dependent particle number is obtaind by the transform in (\ref{e:APN3b}).

\subsection{Time Dependent Oscillator and Adiabatic Particle Number}
\label{ss:QHO}

Another common way to define adiabatic particle number is through the solution to the time-dependent oscillator problem, for each momentum mode $k$ \cite{popov,DittrichReuter,padma,vacha}.
We consider Schwinger vacuum pair production via the Schr\"odinger Picture time evolution of an infinite collection of time-dependent quantum harmonic oscillators, in the presence of a time-dependent  background.
The  sQED hamiltonian becomes
\begin{align}
\hat{H}(t) = \sum_k \left( \frac{1}{2} p_k^2 + \frac{1}{2} \omega_k^2(t) q_k^2 \right)
\label{e:Ham}
\end{align}
where $k$ labels each independent spatial momentum mode, and the field operators map to their quantum mechanical counterparts as $\phi_k \to q_k$ and $\pi_k \to p_k$.
The exact solution of the corresponding time-dependent Schr\"odinger equation can be written as \cite{DittrichReuter,husimi,LewisInvariant1}
\begin{align}
\psi(q_k,t) = \sum_n c_{n,k}(t) \psi_{n}(q_k,t)
\end{align}
where 
\begin{align}
\psi_{n}(q_k,t) = \frac{1}{\sqrt{2^n n!}} \left( \frac{1}{2 \pi \xi_k^2(t)} \right)^{1/4} e^{- \frac{1}{2} \Omega_k(t) q_k^2} H_n\! \left( \frac{q_k}{\sqrt{2}\xi_k(t)} \right) e^{- i (n + \frac{1}{2}) \lambda_{k}(t)}
\label{e:EEF}
\end{align}
Here $\xi_k(t)$ is the solution to the Ermakov-Milne equation (\ref{e:Erm}),  
$\lambda_k(t)$ is defined by (\ref{e:lk}), and the time-dependent function $\Omega_k(t)$ in the Gaussian factor is defined as
\begin{align}
\Omega_k(t) = - i \frac{\dot{\xi_k}}{\xi_k} + \frac{1}{2 \xi_k^2} \;\; . 
\label{e:OXik}
\end{align}
These $\psi_n(q_k, t)$ are normalized eigenfunctions of the exact invariant operator 
\begin{align}
\hat{I}_k(t) = q_k^2\left(\dot{\xi}_k^2+\frac{1}{4 \xi_k^2}\right) +  \xi_k^2 p_k^2 
- \xi_k\, \dot{\xi}_k\left(p_k  q_k+q_k p_k \right)
\end{align}
satisfying
\begin{eqnarray}
\frac{\partial \hat{I}_k}{\partial t}+i\, [\hat{H}, \hat{I}_k]=0
\label{e:inv}
\end{eqnarray}
and
\begin{eqnarray}
\hat{I}_k(t)  \, \psi_n(q_k, t)=\left(n+\frac{1}{2}\right) \, \psi_n(q_k, t)
\end{eqnarray}
The function $\Omega_k(t)$ in (\ref{e:OXik})  is directly related to the Riccati formalism of Section \ref{ss:Rk}, and the mode decomposition of the operator $q_k$, the analog of the field (\ref{e:phik}), in the Heisenberg picture:
\begin{align}
i \Omega_k(t) = \frac{\dot{\xi}_k}{\xi_k} + \frac{i}{2 \xi_k^2} = \frac{\dot{f}_k^*}{f_k^*} = i W_k \left( \frac{1 - r_k^*}{1 + r_k^*} \right) + V_k
\label{e:conn}
\end{align}
Here, $\xi_k(t)$ is again the solution to the Ermakov-Milne equation (\ref{e:Erm}), $f_k(t)$ is the solution to the Klein-Gordon equation (\ref{e:EOM}), and 
the function $r_k(t)$ is related to the reflection amplitude (\ref{e:rk}) by an extra phase:
\begin{align}
r_k(t) = R_k(t) e^{2 i \int^t W_k(t)}
\end{align}
Note that solving for $r_k^*(t)$ in (\ref{e:conn}) in terms of $\xi_k$ leads directly to the analytical form (\ref{e:RK}) of the Riccati reflection probability.

We now define the adiabatic particle number by projecting these states onto a basis set of adiabatically evolving eigenstates of the time-dependent Hamiltonian.
The most general expression for the adiabatically evolving eigenfunction $\zeta_{n}(q_k,t)$, motivated by the assumption of a slowly varying potential given by $\omega_k(t)$, 
takes the form
\begin{align}
\zeta_n(q_k,t) = \frac{1}{\sqrt{2^n n!}} \left( \frac{W_k(t)}{\pi} \right)^{1/4} e^{\tfrac{i}{2} Q_k^*(t) q_k^2} H_n \! \left( \sqrt{W_k(t)} q_k \right)
e^{- i (n + \frac{1}{2}) \int^t W_k(t)}
\label{e:AEF}
\end{align}
where $W_k(t)$ and $V_k(t)$ are basis functions, with the function $Q_k(t)$ defined  as in (\ref{e:Qk}).

At asymptotic early and late times, these adiabatic eigenfunctions reduce to well-defined stationary harmonic oscillator eigenfunctions
\begin{align}
\zeta_n(q_k, t \to \pm \infty) \sim \frac{1}{\sqrt{2^n n!}} \left( \frac{\omega_k( \pm \infty)}{\pi} \right)^{1/4} e^{-\tfrac{1}{2} \omega_k(\pm \infty) q_k^2} H_n \! \left( \sqrt{\omega_k(\pm \infty)} q_k \right) e^{- i (n + \frac{1}{2}) \omega_k(\pm \infty) t}\quad
\label{e:Zn}
\end{align}
A state initially prepared at a particular time can evolve to become a superposition of a variety of states at a later time $t$. 
Assuming that the system is prepared in the ground state at $t = - \infty$, the probability amplitude of making a transition to the $n$-th state is obtained by projecting the adiabatic eigenfunctions $\zeta_n(q_k,t)$ (\ref{e:Zn}) onto the exact eigenfunction (\ref{e:EEF}) for the ground state $\psi_0(q_k,t)$. 
The transition amplitude is 
\begin{align}
C_{n0,k}(t) &= \int^\infty_{- \infty} dq_k \; \zeta_n^*(q_k,t) \psi_0(q_k,t) 
= \left( \frac{1}{2 \xi_k^2 W_k} \right)^{1/4} \sqrt{\frac{2 W_k}{J_k}} \left( \frac{2 W_k}{J_k} - 1  \right)^{n/2} e^{i(n+1/2) \int^t W_k -i \lambda_k(t)}
\label{e:TAn0}
\end{align}
where $n = 0, 2, 4, \dots$. Here $J_k(t) \equiv \Omega_k(t) + i Q_k(t)$. Recalling the form of $\Omega_k(t)$ (\ref{e:OXik}) and $Q_k(t)$ (\ref{e:Qk}), the function $J_k(t)$ simplifies to
\begin{align}
J_k(t) = \left( \frac{1}{2 \xi_k^2} + W_k \right) - i \left( \frac{\dot{\xi}_k}{\xi_k} - V_k \right)
\end{align}
Its modulus squared is related to the Bogoliubov coefficient $\alpha_k(t)$ and the Riccati  reflection probability (\ref{e:RK})
as
\begin{align}
\left| J_k(t) \right|^2 = \frac{2 W_k}{\xi_k^2} \left| \alpha_k(t) \right|^2 = \frac{2 W_k}{\xi_k^2} \left( \frac{1}{1 - \left| R_k \right|^2} \right)
\end{align}
Using this result, the term $\left( \frac{2 W_k}{J_k} - 1 \right)$ in equation (\ref{e:TAn0}) simplifies to 
\begin{align}
\frac{2 W_k}{J_k} - 1 = -
 \frac{\left( \frac{1}{2 \xi_k^2} - W_k \right) - i \left( \frac{\dot{\xi}_k}{\xi_k} - V_k \right)}{\left( \frac{1}{2 \xi_k^2} + W_k \right) - i \left( \frac{\dot{\xi}_k}{\xi_k} - V_k \right)} = r_k^* = R_k^* e^{- 2 i \int^t W_k} \;\; .
\end{align}
Its magnitude is equal to the magnitude of the reflection amplitude $R_k(t)$. Thus the final form for the transition probability from the ground state to the $n$-th state, can be expressed in terms of the reflection probability as 
\begin{align}
\left| C_{n0,k}(t) \right|^2 
= \frac{(n - 1)!!}{n!!} \sqrt{1 - \left| R_k \right|^2} \left| R_k \right|^n,
\label{e:TP}
\end{align}
for $n = 0,2,4,\dots$. For example, the time-dependent vacuum persistence probability, the probability of the system  occupying the ground state at time $t$ is
\begin{align}
\left| C_{00,k}(t) \right|^2 = \sqrt{1 - \left| R_k \right|^2} = \left| \alpha_k(t) \right|^{-1} \;\; .
\end{align}
as expected.

The vacuum expectation value of the state occupation number for a system that adiabatically evolves from being initially prepared in the ground state at $t = - \infty$ is the weighted sum of the transition probabilities (\ref{e:TP}). Utilizing the $R_k$-representation for convenience, the sum simplifies to 
\begin{align}
\tilde{\mathcal{N}}_k(t) &= \sum_{n=0,2,4, \dots}^\infty n \left| C_{n0,k}(t) \right|^2 = \sqrt{1 - \left| R_k \right|^2} \sum_{n=0,2,4, \dots}^\infty \frac{(n-1)!!}{(n-2)!!} \left| R_k \right|^n 
= \frac{\left| R_k \right|^2}{1 - \left| R_k \right|^2} 
\\
	&= \frac{\xi_k^2}{2 W_k} \left[ \left( \frac{1}{2 \xi_k^2} - W_k \right)^2 + \left( \frac{\dot{\xi}_k}{\xi_k} - V_k \right)^2 \right]
	\label{e:APNQ}
\end{align}
Thus, we find exactly the same expression as before, in the Bogoliubov, Riccati and Spectral Function approaches to adiabatic particle number. In the Schr\"odinger picture approach the basis dependence enters through the arbitrariness in (\ref{e:AEF}) of specifying the adiabatic eigenstates $\zeta_n(q_k, t)$ of the time-dependent Hamiltonian.

\section{Adiabatic Expansion and Optimal Adiabatic Approximation of Particle Number}
\label{s:SAPN}

In the preceding section the expression (\ref{e:APNB}) for the  time evolution of the adiabatic particle number,  was equivalently derived through the Bogoliubov, Riccati, Spectral Function and Schr\"odinger approaches.  However, the adiabatic reference mode functions $(W_k,V_k)$ were left unspecified, and the arbitrariness of defining positive and negative energy states implies that an infinite number of consistent choices could be made. In this Section we characterize the different basis choices and identify an optimal one corresponding to the optimal truncation of the adiabatic expansion of (\ref{e:Wcon}). This section also explores the structure and final form of the adiabatic particle number (\ref{e:APNB}), to demonstrate that particle production can be viewed as a measure of small deviations between the exact solution of the Ermakov-Milne equation and various orders of the adiabatic expansion; the deviations from the exact mode function (\ref{e:xik}) by its adiabatic approximation, the reference mode function (\ref{e:refm}) in the Heisenberg formulation, or, equivalently, the deviations from the exact eigenfunction by its adiabatic approximation (\ref{e:AEF}) in the Schr\"odinger formulation.

\subsection{Adiabatic Expansion and Basis Selection}
\label{ss:aexp}

In Section \ref{ss:BTA} we introduced adiabaticity and specified approximate reference mode functions (\ref{e:refm}),
which led to a definition of the time-dependent adiabatic particle number that is dependent on the choice of basis (\ref{e:APNB}). 
We now study and characterize the basis choices.

Insisting that the reference mode functions (\ref{e:refm}) be a solution to the Klein-Gordon equation (\ref{e:EOM}) requires that the function $W_k(t)$ satisfy equation (\ref{e:Wcon}). This equation can be solved by an  adiabatic expansion \cite{BerryAsymptotic,DabrowskiDunne:2014}, in which the leading order is the standard leading WKB solution to (\ref{e:EOM}). This adiabatic expansion is divergent and asymptotic. Successive orders of the adiabatic expansion of $W_k(t)$ are obtained by expanding (\ref{e:Wcon}) in time-derivatives and truncating the expansion at a certain order of derivatives of the fundamental frequency $\omega_k(t)$ (\ref{e:omega}). The up-to-$(j+1)$th order expansion of $W_k(t)$, with the superscript $(j)$ denoting the order of the adiabatic expansion, is then generated by the iterative expansion of
\begin{align}
W_k^{(j+1)}(t) = \sqrt{\omega_k^2(t) - \left[ \frac{\ddot{W}_k^{(j)}(t)}{2 W_k^{(j)}(t)} - \frac{3}{4} \left( \frac{\dot{W}_k^{(j)}(t)}{W_k^{(j)}(t)} \right)^2 \right]}
\label{e:aw}
\end{align}
truncated at terms involving at most $2(j+1)$ derivatives with respect to $t$. 
For the first three orders see \cite{DabrowskiDunne:2014}.
For backgrounds that become constant at asymptotic times it follows that $W_k^{(j)}(\pm \infty) = \omega_k(\pm \infty)$, and $\dot{W}_k^{(j)}(\pm\infty) = 0$ for all $j$.

Despite ambiguity in its explicit form at intermediate times, a critical feature of the real time-dependent function $V_k(t)$ is the necessary requirement that it vanish at asymptotic times
\begin{align}
V_k(\pm \infty) = 0 \;\; .
\label{e:Vinf}
\end{align}
At  asymptotically late time,  the function $V_k$ is no longer ambiguous since the background becomes constant and the identification of particles and anti-particles becomes exact.
In terms of the time-dependent adiabatic particle number (\ref{e:APNB}), this implies that the particle number at future infinity is independent of the choice of $V_k$ (as well as $W_k$). 
At intermediate times, however, the choice critically influences the time evolution of the adiabatic particle number. In the Schr\"odinger approach the basis function $V_k$ is identified from (\ref{e:AEF}) as an unphysical time-dependent phase.
This is equivalently observed in the Bogoliubov, Riccati, and Spectral Function formalisms through the Wronskian condition (\ref{e:Wronk}) where the normalization of the mode function is unaffected by the inclusion of the function $V_k$ in the mode decompositions (\ref{e:fk},\ref{e:dfk}), and thus admits the same interpretation as purely a time-dependent phase.

\begin{figure}[!ht]
\includegraphics[width=\textwidth]{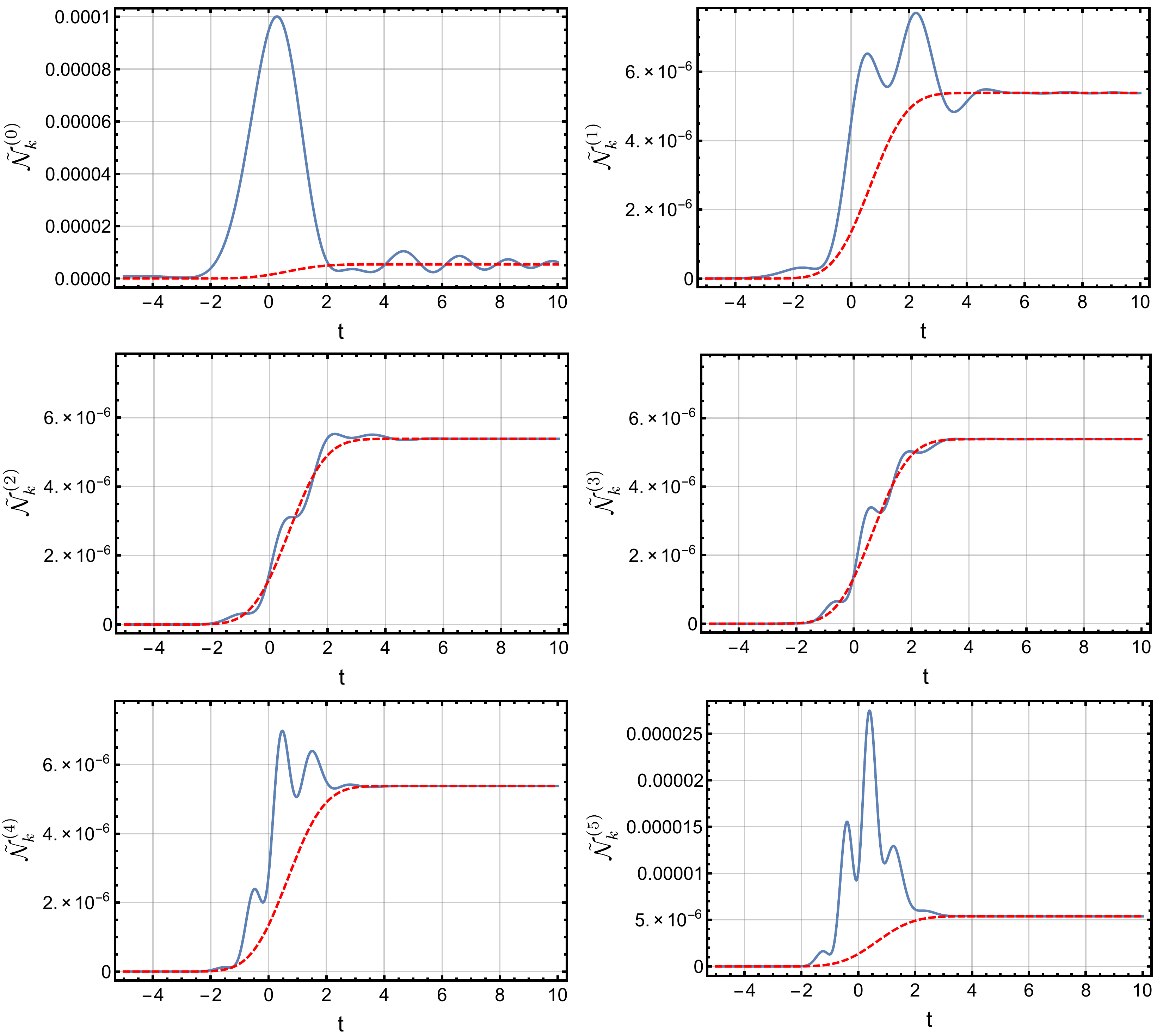}
\caption{The time evolution of the adiabatic particle number (\ref{e:APNJ}) for the first six orders or the adiabatic expansion, considering a time-dependent electric field given by $E(t) = E_0 \text{sech}^2(a t)$ with $E_0=0.25, a=0.1$, longitudinal momentum $k_\parallel = 0$, and transverse momentum $k_\perp = 0$, in units with $m = 1$. The adiabatic particle number (blue-solid line) was computed from (\ref{e:APNJ})  using the exact solution $\xi_k(t)$  to the Ermakov-Milne equation (\ref{e:Erm}) and the $j^{\rm th}$ order adiabatic expansion for the reference function $W^{(j)}_k(t)$ from (\ref{e:aw}). Berry's universal form (\ref{e:bu}) for the particle number is plotted as a red-dashed line. 
We see clearly the typical behavior of an asymptotic expansion, which initally tends towards the optimally truncated form, but then the inclusion of further terms leads to deviation away from this form. For these parameters, the optimal truncation order is $j=3$. 
Notice that the final asymptotic value of the adiabatic particle number is independent of the order of truncation, but that the intermediate time oscillations of the particle number in the conventional  leading order approximation are a factor of 20 times larger than the final value.
}
\label{f:APN1P}
\end{figure}

In this paper we argue that the choice
\begin{align}
V_k(t) \equiv - \frac{\dot{W}_k(t)}{2 W_k(t)}
\label{e:Vk}
\end{align}
is the most suitable and `natural' form.
In the Bogoliubov, Riccati, and Spectral Function approaches the choice (\ref{e:Vk}) arises in the specified mode function decomposition (\ref{e:dfk}) by retaining the contribution from the time-derivative of the $1/\sqrt{2W_k}$ factor in the definition of the reference mode function (\ref{e:refm}).
In the Schr\"odinger approach, the choice (\ref{e:Vk}) appears from insisting that the general form of the adiabatically evolving eigenfunction (\ref{e:AEF}) be a solution to the time-dependent 
Schr\"odinger equation. 
It is a solution provided that the basis function $V_k(t)$ has the form (\ref{e:Vk}), and  yields the same condition on the function $W_k$ as (\ref{e:Wcon}), consistent with the Bogoliubov, Riccati, and Spectral Function approaches.
We adopt this `natural' choice  (\ref{e:Vk}) for the remainder of this paper. In the next Section we explore the dependence on the choice of $W_k(t)$.

\subsection{Optimal Adiabatic Approximation of Particle Number}

Now we consider the specification of the function $W_k(t)$, via various orders of expansion of the adiabatic expansion (\ref{e:aw}). The time evolution of the adiabatic particle number at $j$-th adiabatic order is obtained by inserting the preferred basis (\ref{e:Vk}) into (\ref{e:APNB}) and 
setting $W_k(t) = W_k^{(j)}(t)$ throughout the expression.
At the $j$-th adiabatic order, it takes the form
\begin{align}
\tilde{\mathcal{N}}_k^{(j)}(t) = \frac{\xi_k^2(t)}{2 W_k^{(j)}(t)} \left[ \left( \frac{1}{2 \xi_k^2(t)} - W_k^{(j)}(t) \right)^2 + \left( \frac{\dot{\xi}_k(t)}{\xi_k(t)} + \frac{\dot{W}_k^{(j)}(t)}{2 W_k^{(j)}(t)} \right)^2  \right]
\label{e:APNJ}
\end{align}
%
%
This is completely characterized by the amplitude function $\xi_k(t)$ and the basis function $W_k^{(j)}(t)$.
The general procedure to evaluate (\ref{e:APNJ}) is the following: solve the Klein-Gordon equation (\ref{e:EOM}), or equivalently the Ermakov-Milne equation (\ref{e:Erm}), to obtain $\xi_k(t)$, and compute $W_k^{(j)}(t)$ from the truncation of the adiabatic expansion at the desired adiabatic order. 

The typical behavior of the time evolution of the adiabatic particle is shown in various Figures in this section. We use the explicit example of the single-pulse electric field, for which an analytical solution to $\xi_k$ is known (see Appendix \ref{s:Ex1P}) and the evolution of the adiabatic particle number can thus be analytically obtained. Figure \ref{f:APN1P} illustrates this for the first six orders of the adiabatic expansion. The main observations are:
\begin{enumerate}
\item
Truncating the adiabatic expansion at different adiabatic orders does not affect the final $t=+\infty$  value of the particle number.
\item
Truncating the adiabatic expansion at different adiabatic orders does significantly affect the adiabatic  particle number at intermediate times, in particular in the vicinity of the time of the applied pulse.
\item
Truncating the adiabatic expansion at the optimal  order leads to the smoothest time evolution, which agrees well with the universal form (\ref{e:bu})  found by Berry \cite{BerryAsymptotic,DabrowskiDunne:2014}. 
\item
Going beyond the optimal order again leads to large oscillations in the time vicinity of the applied pulse. This behavior is characteristic of an asymptotic expansion.
\end{enumerate}
A sequential adiabatic order-by-order comparison of the adiabatic particle number in Figure \ref{f:APN1P} shows the typical trend: at intermediate times the adiabatic particle number initially exhibits large oscillations, which become smaller as the optimal order is reached (here, $j=3$), and then increase beyond this optimal order of truncation. This behavior is generic for (divergent and asymptotic) adiabatic expansions where the optimal order of truncation corresponds to a minimum error approximation, and is strongly dependent on the magnitude of the expansion parameter and the parameters found in the effective frequency $\omega_k(t)$ (\ref{e:omega}). 
Dingle found a universal large-order behavior to the adiabatic expansion \cite{dingle}, which was then used by Berry to obtain an approximate universal form to the evolution of the Bogoliubov coefficient $\beta_k(t)$ across a Stokes line when the adiabatic expansion is truncated at optimal order \cite{BerryAsymptotic}. In \cite{DabrowskiDunne:2014} this result was applied to the problem of particle production, leading to the simple universal expression for a single-pulse perturbation
\begin{eqnarray}
\tilde{\mathcal N}_k(t)\approx \frac{1}{4} \left|{\rm Erfc}\left(-\sigma_k(t)\right) e^{- \, F_k^{(0)}} \right|^2
\label{e:bu}
\end{eqnarray}
where the exponential factor $e^{- F_k^{(0)}}$ is determined by the (real-valued and positive) singulant between the complex conjugate pair of turning points 
\begin{eqnarray}
F_k^{(0)} = i \int_{t_c}^{t_c^*} \omega_k(t) \, dt \quad,
\label{f0}
\end{eqnarray}
($t_c$ is the solution of $\omega_k(t_c) = 0$ that is closest to the real axis and located in the upper half plane). The time dependence is given by a universal error function form with argument
\begin{eqnarray}
\sigma_k(t)\equiv \frac{{\rm Im}\, F_k(t)}{\sqrt{2\, {\rm Re}\, F_k(t)}}
\label{sigma}
\end{eqnarray}
where the ``singulant'' function $F_k(t)$ is defined as 
\begin{eqnarray}
F_k(t) = 2i \int^t_{t_c} \omega_k(t^\prime) \, dt^\prime 
\label{sing1}
\end{eqnarray}
For a generalization for multi-pulse perturbations, incorporating quantum interference effects, see \cite{DabrowskiDunne:2014}.

The approximate universal form (\ref{e:bu}) is plotted as a dashed-red curve in Figure \ref{f:APN1P}, showing good agreement with the truncation at the optimal order. 
Figure \ref{f:APN1P} also confirms that the new expression (\ref{e:APNJ})  for the adiabatic particle number agrees with the same adiabatic order-by-order evaluation, obtained numerically, in \cite{DabrowskiDunne:2014}. In comparison to evaluating the coupled time evolution equations (\ref{e:ABevo}) for the Bogoliubov coefficient $B_k^{(j)}(t)$, or the Riccati equation (\ref{e:Rkeq}) for the reflection amplitude $R_k^{(j)}(t)$, a numerical advantage of (\ref{e:APNJ}) is that one does not need to repeatedly solve complicated differential equations for the adiabatic particle number in which the difficulty only increases with truncating the adiabatic expansion at higher orders.
The Klein-Gordon equation (\ref{e:EOM}), or the Ermakov-Milne equation (\ref{e:Erm}), is solved once for $\xi_k(t)$, and then repeatedly projected against different $W_k^{(j)}(t)$, reflecting the truncation of the adiabatic expansion at different adiabatic orders.

\subsection{Particle Production As a Measure of Small Deviations}

In this Section we ``zoom in'' and study the fine details of the time dependence of the particle number. Truncating the adiabatic expansion at different orders  typically has only a small effect on $W_k(t)$, compared to the leading order of the expansion $W_k = \omega_k(t)$, but nonetheless have a large and non-trivial effect on the time evolution of the adiabatic particle number. In this section we explore how these small deviations of the adiabatic approximation influence the evolution of the adiabatic particle number, and show how this indicates the physical phenomenon of quantum interference.

\begin{figure}[!ht]
\centering
\includegraphics[width=\textwidth]{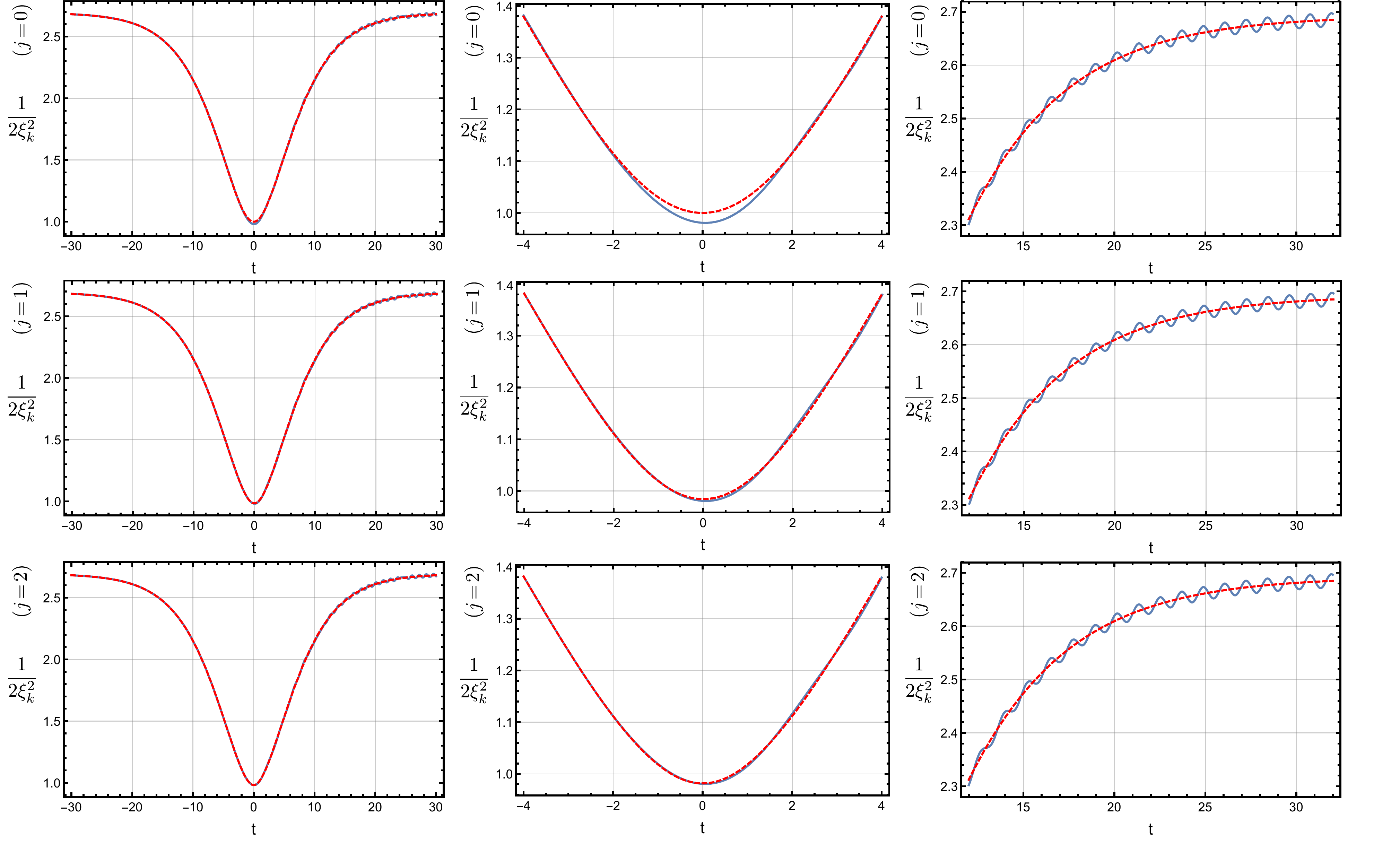}
\caption{
The adiabatic approximation (\ref{e:Xsim}) of $1/(2\xi_k^2)$ (blue-solid line) compared to the adiabatic expansion function $W_k^{(j)}$ (red-dashed line), for the first three orders of the adiabatic expansion, considering a time-dependent single-pulse electric field given by $E_\parallel(t) = E_0 \text{sech}^2(at)$ with the magnitude $E_0 = 0.25$, $a = 0.1$, longitudinal momentum $k_\parallel = 0$, and transverse momentum $k_\perp = 0$, in units with $m = 1$. The central panels zoom in on time-scales near the pulse, while the right-hand panels zoom in on the late-time behavior. 
Notice that the deviations of the approximation from the exact form are typically very small, capturing well the averaged time dependence except near the the peak of the pulse, and except for tiny oscillations about the average value at late times.}
\label{f:WXi2a}
\end{figure}
\begin{figure}[!t]
\centering
\includegraphics[width=\textwidth]{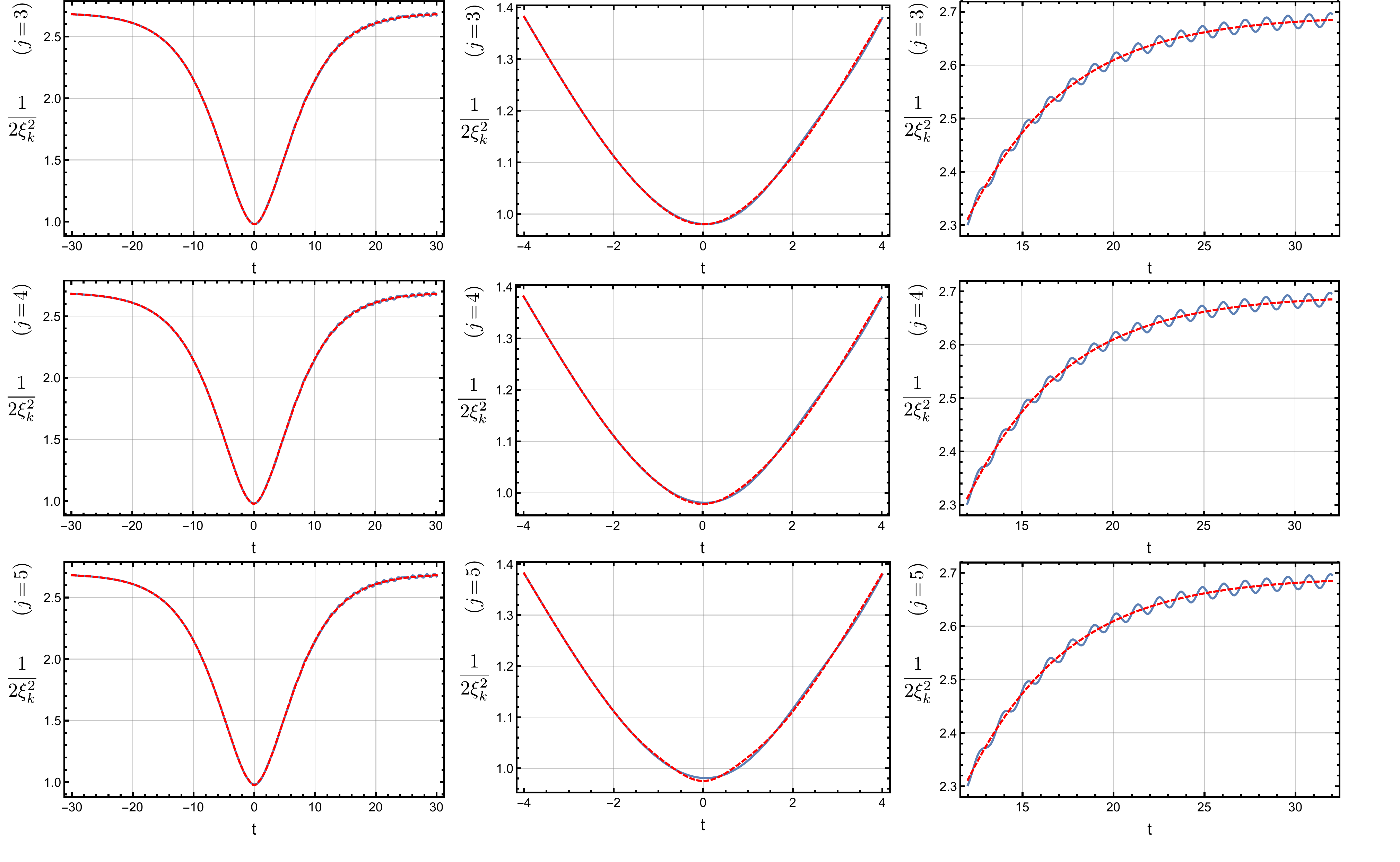}
\caption{
As in Figure  \ref{f:WXi2a} but with plots of the next three adiabatic orders, $j = 3,4,5$ for the adiabatic approximation (\ref{e:Xsim}) of $1/(2\xi_k^2)$ (blue-solid line) compared to the adiabatic expansion function $W_k^{(j)}$ (red-dashed line), considering a time-dependent single-pulse electric field given by $E_\parallel(t) = E_0 \text{sech}^2(at)$ with the magnitude $E_0 = 0.25$, $a = 0.1$, longitudinal momentum $k_\parallel = 0$, and transverse momentum $k_\perp = 0$, in units with $m = 1$. The central panels zoom in on time-scales near the pulse, while the right-hand panels zoom in on the late-time behavior. 
Notice that the deviations of the approximation from the exact form are typically very small, capturing well the averaged time dependence except near the the peak of the pulse, and except for tiny oscillations about the average value at late times.
The optimal order is reached at $j = 3$, after which the deviations begin to grow again. 
}
\label{f:WXi2b}
\end{figure}

\subsubsection{Optimal Adiabatic Approximation of the Ermakov-Milne Equation}
In Section \ref{s:APN}, the adiabatic particle number was found to be determined by the projection of  the solution $f_k(t)$ of the Klein-Gordon equation (\ref{e:EOM}) against a basis set of approximate adiabatic states $\tilde{f}_k(t)$ defined in (\ref{e:refm}). The reference states $\tilde{f}_k(t)$ are chosen to be as good as possible approximations to the exact solution $f_k(t)$, with the appropriate particle production boundary conditions in (\ref{e:Scat}).
Therefore,  the approximation is effectively characterized at the $j^{\rm th}$ order of the adiabatic expansion by
\begin{align}
\xi_k \sim \left(2 W_k^{(j)}\right)^{-\frac{1}{2}}
\label{e:Xsim}
\end{align}
and
\begin{align}
\frac{\dot{\xi}_k}{\xi_k} \sim V_k^{(j)} \equiv - \frac{\dot{W}_k^{(j)}}{2 W_k^{(j)}}
\label{e:dXsim}
\end{align}
This last approximation can be equivalently seen as neglecting the exponentially small $r_k^*$ in (\ref{e:conn}). 
Notice that the approximation (\ref{e:dXsim}) is the ratio form of the first derivative of approximation (\ref{e:Xsim}), and thus is consistent with the `natural' basis choice (\ref{e:Vk}). 

The structure of the adiabatic particle number in (\ref{e:APNJ}) is explicitly composed of the differences of the adiabatic approximations (\ref{e:Xsim}, \ref{e:dXsim}). We now examine how these approximations work in practice with the adiabatic expansion truncated at various adiabatic orders. Figures \ref{f:WXi2a} and \ref{f:WXi2b} examine the adiabatic approximation (\ref{e:Xsim}) by directly comparing $1/(2\xi_k^2)$ with the adiabatic functions $W_k^{(j)}$ for various adiabatic orders, considering a single-pulse time-dependent electric field of the form
\begin{align}
E_\parallel(t) = E_0 \text{sech}^2(at)
\label{e:E1P}
\end{align}
given by the time-dependent vector potential
\begin{align}
A_\parallel(t) = - \frac{E_0}{a} \tanh(at) 
\label{e:A1P}
\end{align}
\begin{figure}[!ht]
\centering
\includegraphics[width=\textwidth]{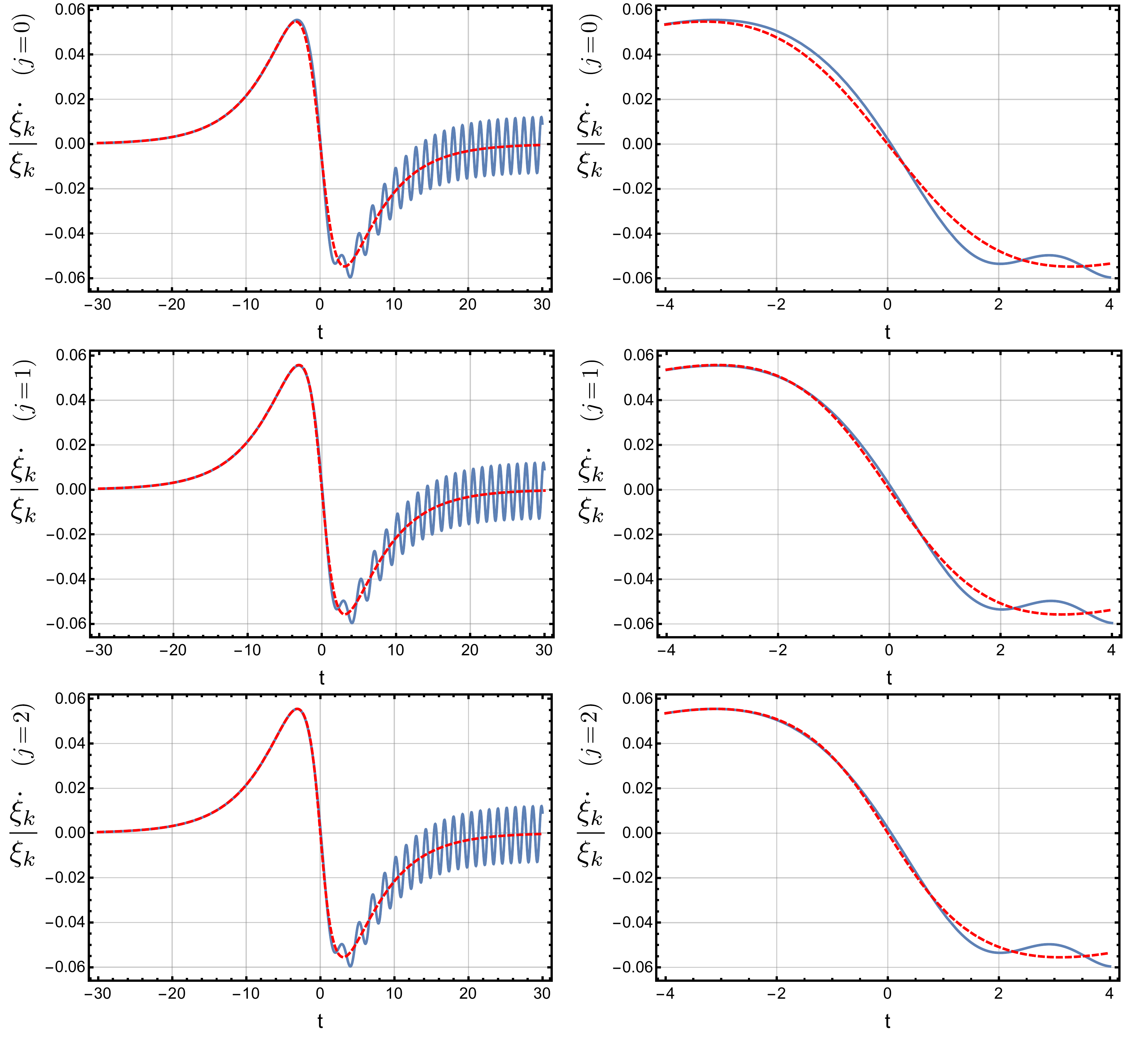}
\caption{
Plots of approximation (\ref{e:dXsim}), the time derivative form of (\ref{e:Xsim}) as a ratio, by comparing $\dot{\xi}_k / \xi_k$ (blue-solid line) with the adiabatic function $V_k^{(j)} = - \dot{W}_k^{(j)} / (2W_k^{(j)})$ (red-dashed line), for the first three orders of the adiabatic expansion, considering a time-dependent single-pulse electric field given by $E_\parallel(t) = E_0 \text{sech}^2(at)$ with the same pulse parameters as for Figure \ref{f:WXi2a}. The left-hand panels show the time-evolution over a wide range of $t$, and the right-hand panels zoom in on the vicinity of the pulse.
The approximate (red-dashed) curves accurately describe the averaged time evolution, but miss the late-time oscillations, which are more pronounced than those in Figures \ref{f:WXi2a} and \ref{f:WXi2b} because they are effectively the derivatives of those small oscillations.}
\label{f:dWXi2a}
\end{figure}
\begin{figure}[!ht]
\centering
\includegraphics[width=\textwidth]{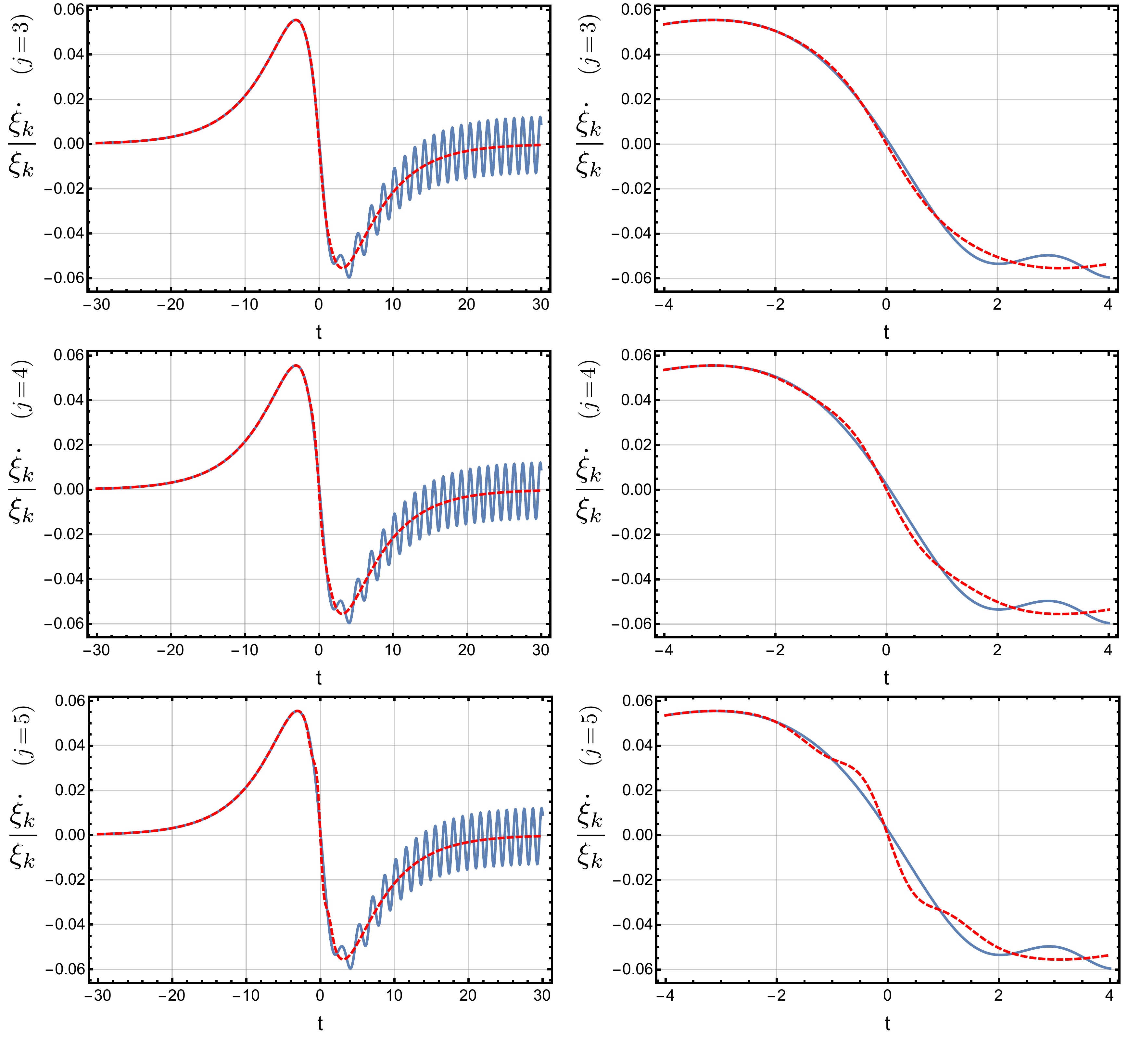}
\caption{ 
As in Figure \ref{f:dWXi2a}, but with plots of the next three adiabatic orders, $j = 3,4,5$,  showing the approximation (\ref{e:dXsim}), the time derivative form of (\ref{e:Xsim}) as a ratio, by comparing $\dot{\xi}_k / \xi_k$ (blue-solid line) with the adiabatic function $V_k^{(j)} = - \dot{W}_k^{(j)} / (2W_k^{(j)})$ (red-dashed line), considering a time-dependent single-pulse electric field given by $E_\parallel(t) = E_0 \text{sech}^2(at)$ with the same pulse parameters as for Figure \ref{f:WXi2a}. The left-hand panels show the time-evolution over a wide range of $t$, and the right-hand panels zoom in on the vicinity of the pulse.
The approximate (red-dashed) curves accurately describe the averaged time evolution, but miss the late-time oscillations, which are more pronounced than those in Figures \ref{f:WXi2a} and \ref{f:WXi2b} because they are effectively the derivatives of those small oscillations.
The optimal truncation order is at $j=3$, after which the deviations begin to increase again.
}
\label{f:dWXi2b}
\end{figure}
Figure \ref{f:WXi2a} considers the first three adiabatic orders, while Figure \ref{f:WXi2b} considers the next three orders. The left-hand figures show the time-evolution over a wide range of $t$; the central panels zoom in on the vicinity of the pulse, and the right-hand panels zoom in on the late time behavior. 
Notice that the approximation (\ref{e:Xsim}) is extremely good, with only very small deviations between $W_k^{(j)}(t)$ and $1/(2\xi_k^2(t))$, which moreover do not change in any particularly dramatic fashion as the truncation order changes. Notice the tiny oscillations about an accurate time-averaged approximation at late times. There are small deviations near the time location of the pulse, which shrink until the optimal order and then begin to grow again.
Again, this is typical of adiabatic expansions where the optimal order of truncation corresponds to a minimum error approximation, and results in $W_k^{(j)}$ corresponding to the optimal adiabatic approximation of $1/(2\xi_k^2)$. This optimal approximation represents a simple `best possible' approximation.

Figures \ref{f:dWXi2a} and \ref{f:dWXi2b} examine the adiabatic approximation (\ref{e:dXsim}), the first derivative ratio form of approximation (\ref{e:Xsim}), by directly comparing $\dot{\xi}_k / \xi_k$ with the basis function $V_k^{(j)} = - \dot{W}_k^{(j)} / (2 W_k^{(j)})$, evaluated at  various orders $j$ of the adiabatic expansion, using the same electric field configuration and parameters as used for Figures \ref{f:WXi2a} and \ref{f:WXi2b}. Figure \ref{f:dWXi2a} considers the first three adiabatic orders, while Figure \ref{f:dWXi2b} considers the next three orders. In both Figures  \ref{f:dWXi2a} and \ref{f:dWXi2b}, $\dot{\xi}_k/\xi_k$ is plotted as a solid-blue curve, and $V_k^{(j)} = - \dot{W}_k^{(j)} / (2 W_k^{(j)})$ is plotted as a dashed-red curve. The left-hand panels show the time-evolution over a wide range of $t$, and the right-hand panels zoom in on the vicinity of the pulse. Notice that there are once again  oscillations about an accurate time-averaged approximation at late times, but that these oscillations are now larger than those seen at late times in Figures \ref{f:WXi2a} and \ref{f:WXi2b}. This is because this is effectively measuring the {\it derivatives} of the tiny late-time oscillations in Figures \ref{f:WXi2a} and \ref{f:WXi2b}. We also see that the changes from one order of truncation to the next are not particularly pronounced.
 
\begin{figure}[!ht]
\centering
\includegraphics[width=\textwidth]{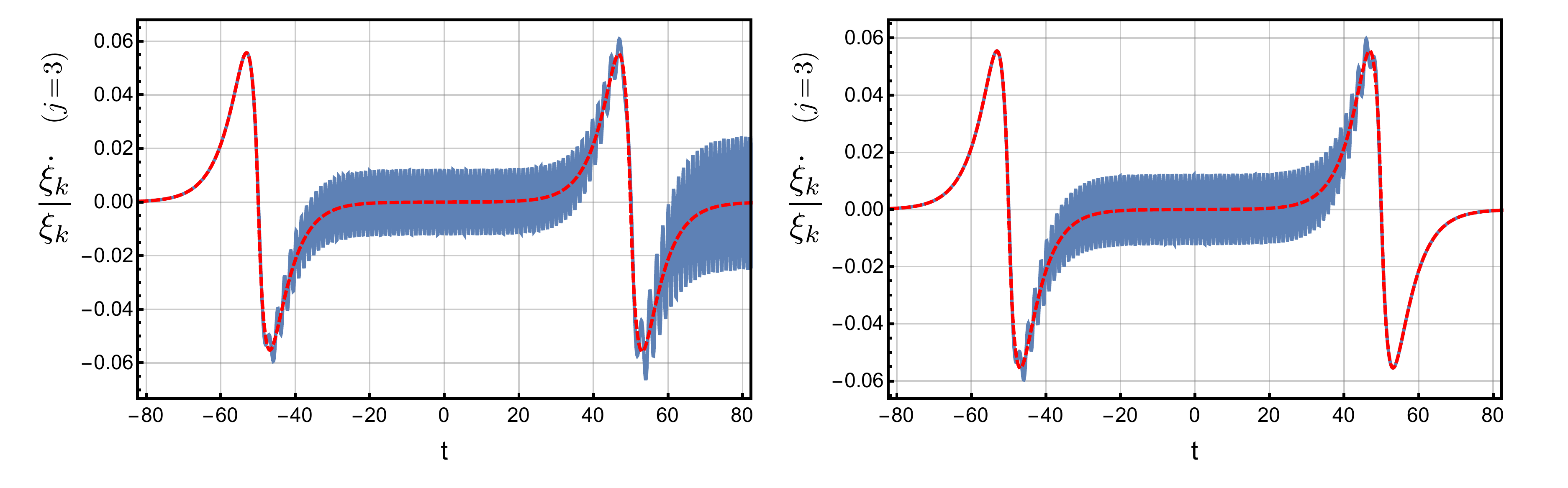}
\caption{
Plots of the approximation (\ref{e:dXsim}) at the optimal truncation order ($j=3$), for the time derivative form of (\ref{e:Xsim}) as a ratio, by comparing $\dot{\xi}_k /\xi_k$ (blue-solid line) and $V_k^{(j)} = -\dot{W}_k^{(j)} / 2W_k^{(j)}$ (red-dashed line), for a time-dependent electric field given by (\ref{e:E2P}) with the magnitude $E_0 = 0.25$, $a = 0.1$, $b=50$, and transverse momentum $k_\perp = 0$, in units with $m = 1$. 
The longitudinal momentum $k_\parallel$ was selected to correspond to maximum constructive ($k_\parallel = 2.51555$ for left subplot), and maximum destructive interference ($k_\parallel = 2.49887$ for right subplot) in the particle number at future infinity \cite{dumludunne,AkkermansDunne:2012}. In the maximum constructive case [left panel], the oscillations introduced after each pulse interfere to increase in magnitude, and double in scale. In the maximum destructive case [right panel], the oscillations introduced by the first and second pulse interfere to completely cancel.
}
\label{f:WXiBoth}
\end{figure}

\begin{figure}[!ht]
\centering
\includegraphics[width=\textwidth]{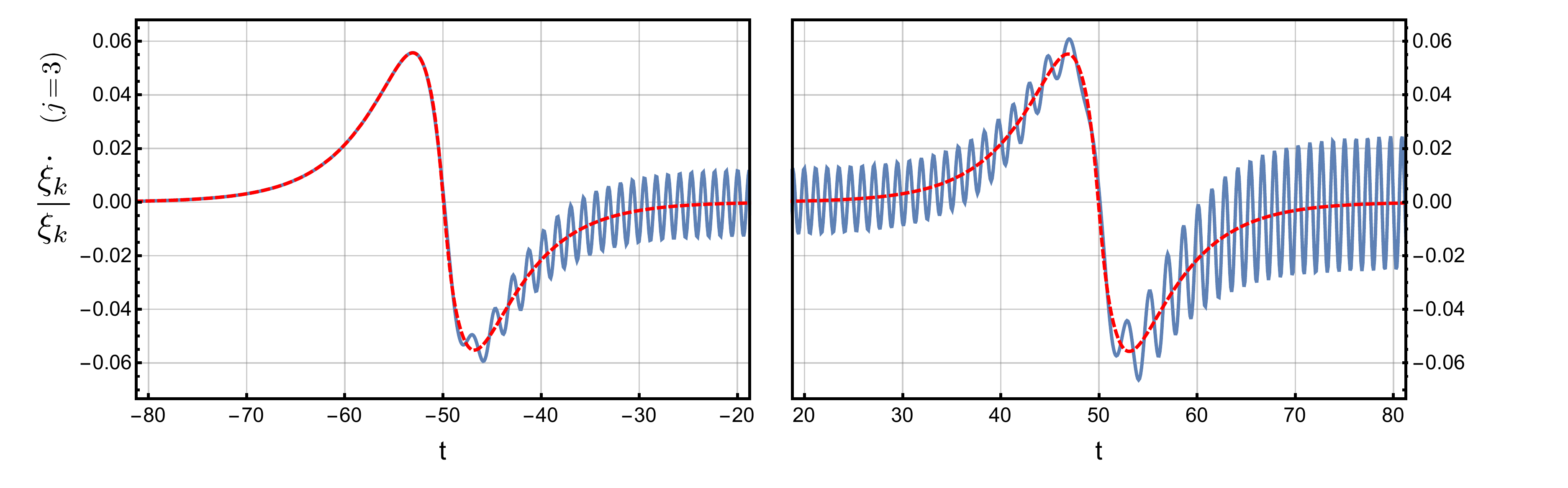}
\caption{
Zoomed-in view of  the left subplot of Figure \ref{f:WXiBoth}, plotted for a closer examination of the approximation (\ref{e:dXsim}) in the vicinity of the pulse centers $(t=\pm50)$, for the case of maximum constructive interference.
}
\label{f:WXi2PC}
\end{figure}

\begin{figure}[!ht]
\centering
\includegraphics[width=\textwidth]{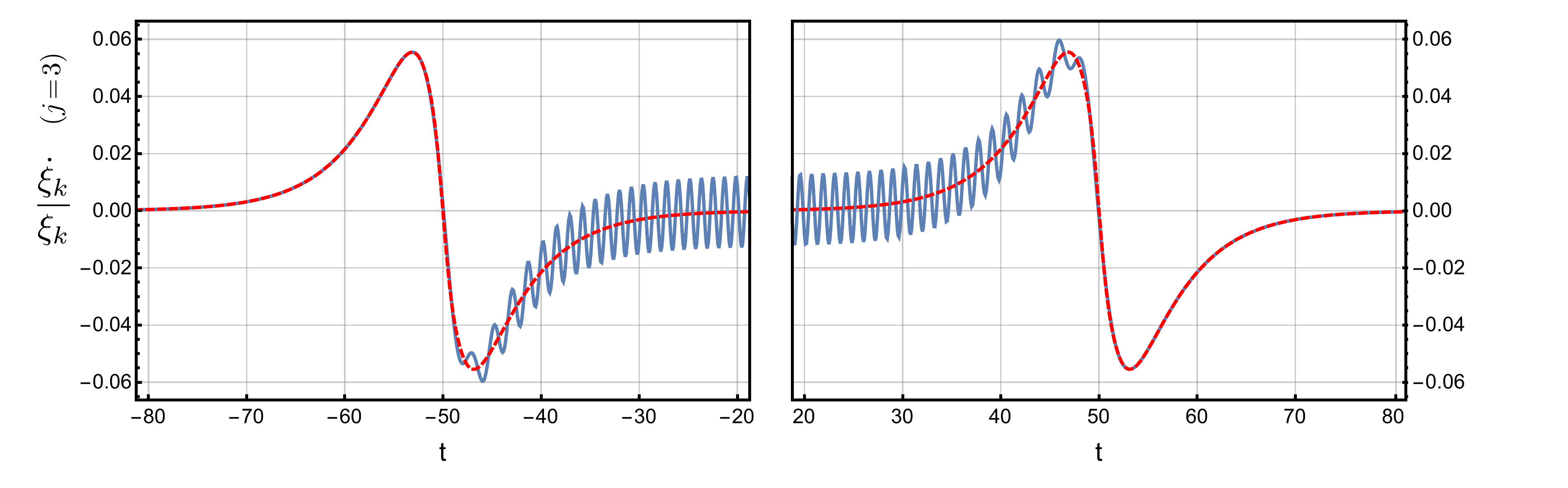}
\caption{
Zoomed-in view of  the right subplot of Figure \ref{f:WXiBoth}, plotted for a closer examination of the approximation (\ref{e:dXsim}) in the vicinity of the pulse centers $(t=\pm50)$, for the case of maximum destructive interference.
}
\label{f:WXi2PD}
\end{figure}

We now examine the approximations by considering a time-dependent electric field with non-trivial temporal structure, to illustrate  the phenomenon of quantum interference.
Figures \ref{f:WXiBoth}, \ref{f:WXi2PC} and \ref{f:WXi2PD} examine the adiabatic approximation (\ref{e:dXsim}) by considering the alternating sign double-pulse electric field of the form
\begin{align}
E_\parallel(t) = - E_0 \, \text{sech}^2\left[ a ( t + b) \right] + E_0 \, \text{sech}^2\left[ a ( t - b) \right]
\label{e:E2P}
\end{align}
given by the time-dependent vector potential
\begin{align}
A_\parallel(t) = - \frac{E_0}{a} \big( - \tanh\left[ a ( t + b) \right] +  \tanh\left[ a ( t - b) \right] \big) .
\label{e:A2P}
\end{align}
Figure \ref{f:WXiBoth} compares the adiabatic approximation (\ref{e:dXsim})  at the optimal order of truncation,  $j=3$, for two different cases of constructive (left panel), and destructive (right panel) interference. At this optimal order, $V_k^{(3)}$ corresponds to the optimal adiabatic approximation of $\dot{\xi}_k/\xi_k$, accurately capturing the average of its amplitude at all times, but missing the oscillatory behavior, which encodes critical information regarding particle production.
Figures~\ref{f:WXi2PC} and \ref{f:WXi2PD} show zoomed-in views, near each of the pulses,   for the left and right panels of Figure \ref{f:WXiBoth}, respectively. Figures~\ref{f:WXi2PC} and \ref{f:WXi2PD} are plotted with the same pulse parameters but with different longitudinal momentum to highlight the manifestation of quantum interference that are associated with electric fields having non-trivial temporal structure \cite{dumludunne,AkkermansDunne:2012}.
Specifically, the longitudinal momentum in Figure~\ref{f:WXi2PC} corresponds to maximum {\it constructive} interference in the adiabatic particle number at asymptotic times, while the longitudinal momentum in Figure~\ref{f:WXi2PD} corresponds to maximum {\it destructive} interference.
A similar adiabatic order-by-order comparison of the adiabatic approximation  shows the same trend observed in Figures \ref{f:WXi2a}, \ref{f:WXi2b}, \ref{f:dWXi2a}, and \ref{f:dWXi2b}: the matching of both sides of approximation (\ref{e:dXsim}) improve until the optimal order is achieved, and then grows more and more mismatched after this optimal order of truncation. The oscillations in $\dot{\xi}_k / \xi_k$ directly correspond to quantum interference: in Figure \ref{f:WXi2PC}, we observe oscillations that increase in magnitude as a result of each pulse and constructively interfere with one another to double in magnitude; while in Figure~\ref{f:WXi2PD} we observe oscillations that increase in magnitude as a result of the first pulse but then cancel completely as the oscillations introduced by the second pulse destructively interfere with the first.
Note that the magnitude of the oscillations in between the two pulses in Figures \ref{f:WXi2PC} and \ref{f:WXi2PD}, which are widely temporally separated, are equal to the magnitude of the oscillations at asymptotic times in the single-pulse case in Figures~\ref{f:dWXi2a} and \ref{f:dWXi2b}.

\subsubsection{Adiabatic Particle Number as a Measure of Small Deviations}

In this subsection we examine how the small deviations from the adiabatic approximations (\ref{e:Xsim}, \ref{e:dXsim}) determine the adiabatic particle number. We re-write the expression (\ref{e:APNJ}) as the sum of two terms, the first of which measures the deviations of the adiabatic approximation (\ref{e:Xsim}), and the second of which measures the deviations of the adiabatic approximation (\ref{e:dXsim}):
\begin{align}
\tilde{\mathcal{N}}_k^{(j)}(t)  = 
\frac{1}{4}  \left(  \sqrt{2 W_k^{(j)}(t) \xi_k^2(t)} - \frac{1}{ \sqrt{2 W_k^{(j)}(t) \xi_k^2(t)}} \right)^2  +\frac{1}{4}
\left( \frac{\frac{d}{dt}\left(\sqrt{2 W_k^{(j)}(t) \xi_k^2(t)}\right)}{W_k^{(j)}(t)} \right)^2 
\label{e:APNg}
\end{align}

As shown in the previous subsection, the relationship between the exact solution $\xi_k(t)$ to the Ermakov-Milne equation and the adiabatic expansion functions $W_k^{(j)}(t)$ are given by the approximations (\ref{e:Xsim}, \ref{e:dXsim}). The structure of the adiabatic particle number (\ref{e:APNJ}) specifically extracts the very small changes introduced by truncating the adiabatic expansion at different orders and the small oscillations from the exact solution to the Ermakov-Milne equation that directly encode the particle production phenomenon. This yields a new perspective: particle production is characterized by the measure of these small deviations. The first term on the right-hand-side of (\ref{e:APNg}) measures the deviations of $2 W_k^{(j)}(t) \xi_k^2(t)$ from 1, while the second term on the right-hand-side of (\ref{e:APNg}) effectively measures the derivatives of this deviation.

\begin{figure}[!ht]
\centering
\includegraphics[width=\textwidth]{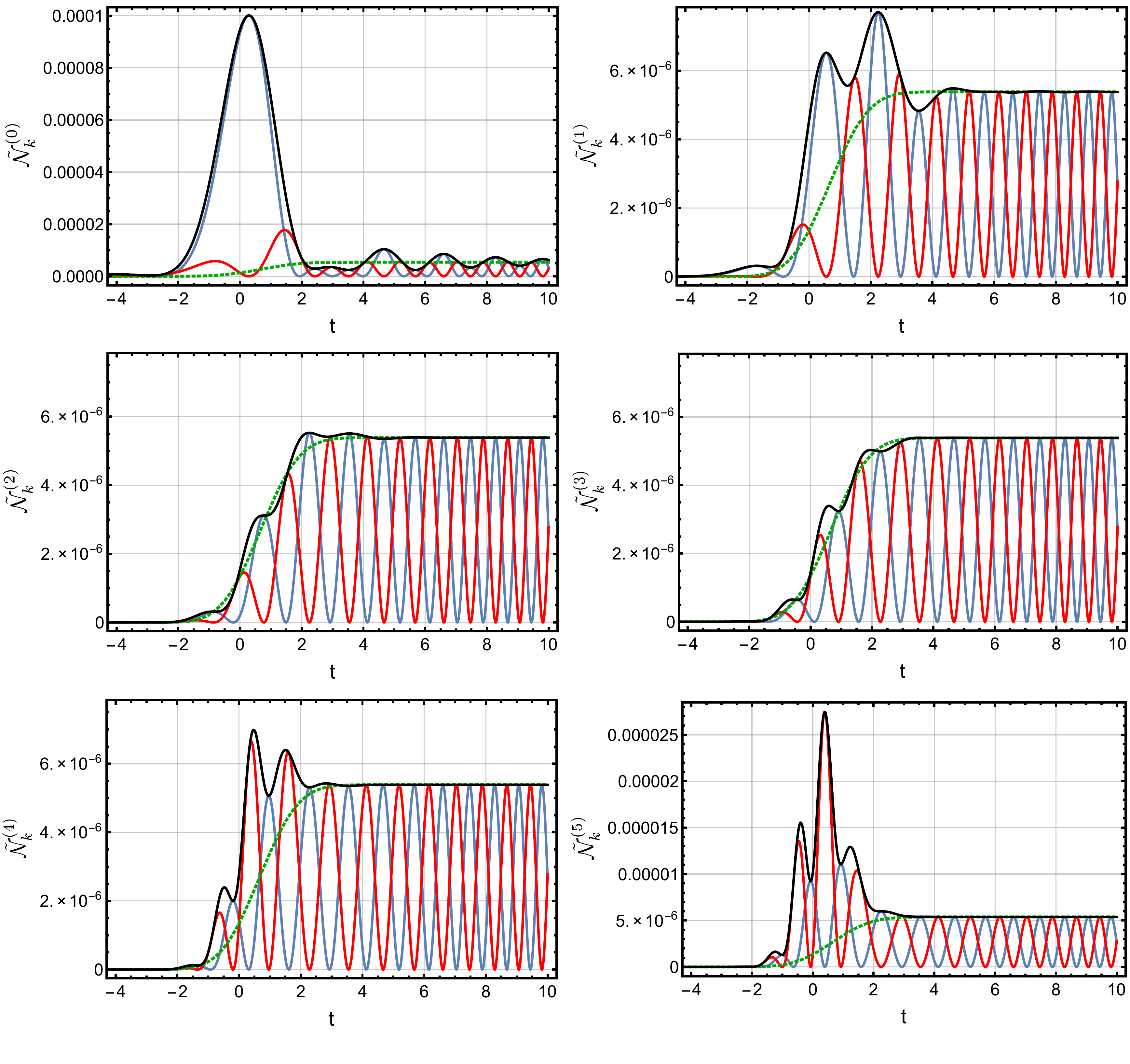}
\caption{
The time evolution of the adiabatic particle number (\ref{e:APNg}) (black-solid line), and its components, the first (blue-solid line) and second (red-solid line) terms on the right-hand-side of (\ref{e:APNg}), for the first six orders of the adiabatic expansion, considering a time-dependent single-pulse electric field given by $E(t) = E_0 \text{sech}^2(at)$ with $E_0 = 0.25, a = 0.1$, longitudinal momentum $k_\parallel = 0$, and transverse momentum $k_\perp$ = 0, in units with $m=1$. 
Notice that each of the components (blue and red curves) is highly oscillatory, and out of phase, especially at late times, but the sum is smooth except in the vicinity of the pulse. Also note the difference in scales in the various sub-plots. The deviations decrease dramatically as the optimal order ($j=3$) is approached, and then grow again as this order is passed. Further features are discussed in the text.
}
\label{f:PN1P}
\end{figure}

In Figure \ref{f:PN1P} we see the results of these small deviations. The black solid line in Figure \ref{f:PN1P} shows the exact adiabatic particle number, for the first six orders of the adiabatic expansion. These are the curves plotted previously as solid blue lines in Figure \ref{f:APN1P}. The blue and red curves in Figure \ref{f:PN1P} show, respectively,  the first and second terms on the right-hand-side of (\ref{e:APNg}). Notice that their combined envelope matches the adiabatic particle number (the black curve), but the blue and red curves oscillate out of phase with one another, since the latter effectively characterizes the time derivative of the former. 
Each component is highly oscillatory, especially at late times, but their envelope is smooth except in the vicinity of the pulse. Also notice the difference of scales in the various sub-plots. The deviations decrease significantly as the optimal order is approached, and then grow again as this order is passed. The green dashed line shows Berry's universal approximation, which matches the optimally truncated order of the adiabatic expansion (here $j=3$). At the optimal order, we see the culmination of the optimal adiabatic approximation of the Ermakov-Milne equation in the final answer of the particle number: the scale of the oscillations of both components become comparable, they level off much more quickly, and sum to yield the smoothest time evolution of the adiabatic particle number.

\begin{figure}[!ht]
\centering
\includegraphics[width=\textwidth]{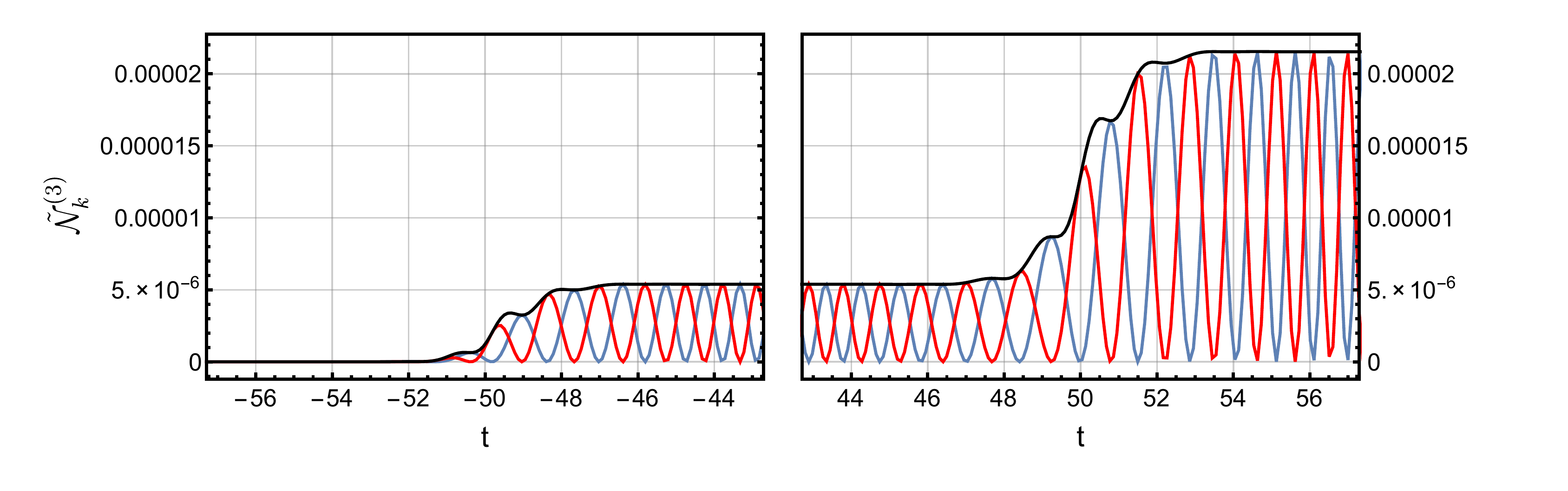}
\caption{ 
The time evolution of the adiabatic particle number (\ref{e:APNg}) (black-solid line), and its components, the first (blue-solid line) and second (red-solid line) terms on the right-hand-side of (\ref{e:APNg}), at the optimal order of the adiabatic expansion, 
for the time-dependent double-pulse electric field given by (\ref{e:E2P}) with $E_0 = 0.25, a = 0.1, b = 50$, longitudinal momentum $k_\parallel = 2.51555$, and transverse momentum $k_\perp$ = 0, in units with $m=1$.
The longitudinal momentum was selected to correspond to maximum constructive interference, with the final value being $4$ times the intermediate plateau value between the two pulses. 
 At intermediate times, notice the phase difference of the oscillatory components of (\ref{e:APNg}), which remarkably sum to a smooth evolution of the adiabatic particle number.}
\label{f:PN2PC}
\end{figure}

\begin{figure}[!ht]
\centering
\includegraphics[width=\textwidth]{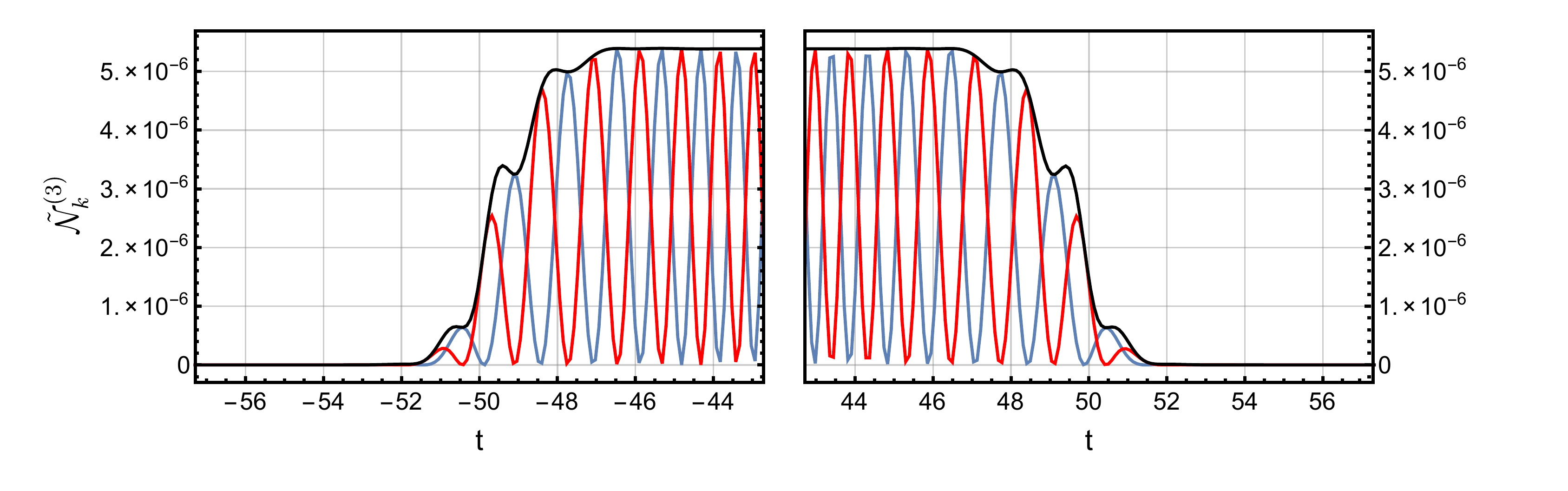}
\caption{ 
The time evolution of the adiabatic particle number (\ref{e:APNg}) (black-solid line), and its components, the first (blue-solid line) and second (red-solid line) terms on the right-hand-side of (\ref{e:APNg}), at the optimal order of the adiabatic expansion, 
for the time-dependent double-pulse electric field given by (\ref{e:E2P}) with $E_0 = 0.25, a = 0.1, b = 50$, longitudinal momentum $k_\parallel =  2.49887$, and transverse momentum $k_\perp$ = 0, in units with $m=1$. The longitudinal momentum was selected to correspond to maximum destructive interference, with vanishing final particle number at future infinity.
 At intermediate times, notice the phase difference of the oscillatory components of (\ref{e:APNg}), which remarkably sum to a smooth evolution of the adiabatic particle number.}
\label{f:PN2PD}
\end{figure}

In Figures \ref{f:PN2PC} and \ref{f:PN2PD} we plot in blue and red the same two components of the right-hand-side of the expression (\ref{e:APNg}), but consider the alternating sign double-pulse electric field given by (\ref{e:E2P}). Both Figures utilize the same pulse parameters, but with different longitudinal momentum that correspond to maximum constructive interference, Figure \ref{f:PN2PC}, and maximum destructive interference, Figure \ref{f:PN2PD}.
The constructive interference can be seen in Figure \ref{f:PN2PC} as the final value of the particle number at future infinity $\tilde{\mathcal{N}}_k(+\infty)$ is $4=2^2$ times the value at times in between the two pulses. The destructive interference can be seen in Figure \ref{f:PN2PD} through  the vanishing final value of the particle number at future infinity $\tilde{\mathcal{N}}_k(+\infty)$. 
Both figures show just  the optimal order of truncation of the adiabatic expansion ($j=3$ for these parameters), but we have confirmed that a similar adiabatic order-by-order comparison shows the same trend exhibited in the phase and scale of the oscillations, as seen in Figure \ref{f:PN1P}. 
In Figure \ref{f:PN2PC}, the interference results in the components being out of phase in such a way to produce enhancement of particle production, which follow an $n^2$ coherence pattern \cite{DabrowskiDunne:2014}, while in Figure \ref{f:PN2PD} it leads to cancellation with no particles produced at the final time. Again, in each case, the two different components of (\ref{e:APNg}) remarkably sum to produce a smooth evolution of the particle number at all times.

\section{Conclusion}
\label{s:Conc}

In this paper we have explored the detailed structure of the time evolution of the adiabatic particle number for particle production in time-dependent electric fields (the Schwinger effect). Through the Ermakov-Milne equation (\ref{e:Erm}), the amplitude of the solution to the Klein-Gordon equation (\ref{e:EOM}), an analytic expression (\ref{e:APNJ}) for the time-dependent adiabatic particle number was derived  by projection of the  exact solution $\xi_k(t)$ against a basis of approximate adiabatic reference states, characterized by the function $W_k(t)$ defined in (\ref{e:refm}), and its various orders of adiabatic approximation defined in (\ref{e:aw}). The form of expression  (\ref{e:APNJ}) clearly illustrates the separation between the exact solution and the choice of adiabatic basis, and  illustrates the role of the adiabatic approximation in defining the reference states. It also simplifies  its numerical evaluation, as $\xi_k(t)$ need only be computed once, independent of the order of truncation of the adiabatic approximation for the reference states. 
We showed that the Bogoliubov, Riccati,  Spectral Function, and  Schr\"odinger approaches to the adiabatic particle number each yield the same analytic expression for the particle number, indicating that this form of basis dependence is a universal feature of the definition of the adiabatic particle number at intermediate times. Note that the final particle number, at $t=+\infty$, is independent of the basis choice, but at intermediate times the particle number is highly sensitive to the basis choice.
A variety of cases were illustrated and the new form (\ref{e:APNJ}) agrees with previously reported numerical  results in \cite{DabrowskiDunne:2014}.

This leads to a proposal for an optimal adiabatic particle number, at all times, even during the time evolution.
The logic is the following. We first showed that the adiabatic particle number at intermediate times is basis dependent, and therefore presumably unphysical since the order of truncation of the asymptotic adiabatic expansion depends sensitively on the physical parameters of the driving perturbation. But the situation is completely reversed  due to the remarkable universality of the smoothing of the Stokes phenomenon found by Berry \cite{BerryAsymptotic}, using which we  argued that it is in fact possible to define an optimal adiabatic particle number. This universality means that no matter what is the optimal order, the time dependence of the optimally truncated particle number will have the same error-function time dependence form in (\ref{e:bu}). This is confirmed through a number of examples. Further physical support for this proposed definition comes from the resulting clear view of quantum interference effects, illustrated here for the double-pulse sign-alternating electric field in (\ref{e:E2P}), which exhibits both constructive and destructive interference, depending on the longitudinal momentum of the produced particles. The structure of (\ref{e:APNJ}) also shows that the adiabatic particle number may be characterized by the small deviations between the exact solution and its adiabatic approximations. 
At the optimal order of truncation of the adiabatic approximation, the deviations are the smallest and smoothest, and correspond to an optimal adiabatic approximation.

Future work will address the implications of these results for back-reaction and non-equilibrium processes \cite{KME:1998,Habib:1999cs,Rau:1995ea,Zahn:2015awa}, in both the Schwinger effect and related time-dependent non-equilibrium systems such as heavy ion collisions and driven multi-level quantum systems.

\bigskip

{\bf Acknowledgement:} This material is based upon work supported by the U.S. Department of Energy, Office of Science, Office of High Energy Physics, under Award Number DE-SC0010339. 

\bigskip

\section{Appendix: Single-pulse Analytical Example}
\label{s:Ex1P}

An analytic example that is commonly used in the literature \cite{Narozhnyi:1970uv} in connection with the adiabatic particle number is the  single-pulse electric field given by (\ref{e:E1P}) with the vector potential (\ref{e:A1P}). 
The solution of the Klein-Gordon equation (\ref{e:EOM}) with this electric field case is a hypergeometric solution of the form
\begin{align}
f_k(t) = \frac{1}{\sqrt{2 \omega_k(-\infty)}} (- x)^{- i \mu} (1 - x)^{\epsilon} {}_{2}F_1\left[ \epsilon - i (\mu + \nu) , \epsilon - i (\mu - \nu) , 1 - 2 i \mu , x \right]
\label{e:anal}
\end{align}
where $x \equiv - e^{2 a t}$, and 
\begin{align}
&
&
\epsilon &= \frac{1}{2} \left( 1 + \sqrt{1 - \frac{4 E_0^2}{a^4}} \right)
&
\mu &= \frac{\omega_k(-\infty)}{2a}
&
\nu = \frac{\omega_k(+\infty)}{2a}
&
&
\end{align}
which satisfies the Wronskian condition (\ref{e:Wronk}) and matches the rightward scattering scenario (\ref{e:Scat}).
From (\ref{e:xik}), then the absolute magnitude of (\ref{e:anal}) is the analytic solution to the Ermakov-Milne equation (\ref{e:Erm}). 
The scattering coefficients in equation (\ref{e:Scat}) with (\ref{e:anal}) are 
\begin{align}
A_k &= \frac{\Gamma\left( 1 - 2 i \mu \right) \Gamma\left( - 2 i \nu \right)}{\Gamma\left( \epsilon - i (\mu + \nu) \right) \Gamma\left( 1 - \epsilon - i (\mu + \nu) \right)} \\
B_k &= \frac{\Gamma\left( 1 - 2 i \mu \right) \Gamma\left( 2 i \nu \right)}{\Gamma\left( \epsilon - i (\mu - \nu) \right) \Gamma\left( 1 - \epsilon - i (\mu - \nu) \right)}
\end{align}
where the coefficients satisfy unitarity, $\left| A_k \right|^2 - \left| B_k \right|^2 = 1$, and are related to the final-time Bogoliubov coefficients by $A_k = \alpha_k(+\infty)$ and $B_k = \beta_k(+\infty)$. Thus, the final particle number at future infinity for the single-pulse case is precisely $\tilde{\mathcal{N}}_k(+\infty) = \left| B_k \right|^2$.


\begin{thebibliography}{999}
%
%
%


\bibitem{he}
W. Heisenberg and H. Euler, ``Consequences of Dirac's Theory of Positrons,'' 
Z. Phys. {\bf 98}, 714 (1936).

\bibitem{sch}
J. Schwinger, ``On gauge invariance and vacuum polarization,''
Phys. Rev. {\bf 82} (1951) 664.

\bibitem{greiner}
W. Greiner, B. M\"uller and J. Rafelski, \emph{Quantum Electrodynamics of Strong Fields}, 
(Springer, Berlin, 1985).

\bibitem{dunne}
G.~V.~Dunne,  ``Heisenberg-Euler effective Lagrangians: Basics and extensions,'' in  Ian Kogan Memorial Collection, {\it 'From Fields to Strings: Circumnavigating Theoretical Physics'} Volume 1,
M.~Shifman et al (ed.), (World Scientific, 2005), 
\hhref{hep-th/0406216}.

\bibitem{parker68}
L. Parker, ``Particle creation in expanding universes,''
Phys. Rev. Lett. {\bf 21}, 562 (1968);
`Quantized fields and particle creation in expanding universes. 1.,''
  Phys.\ Rev.\  {\bf 183}, 1057 (1969);
  ``Quantized fields and particle creation in expanding universes. 2.,''
  Phys.\ Rev.\ D {\bf 3}, 346 (1971)
  [Erratum-ibid.\ D {\bf 3}, 2546 (1971)].

\bibitem{zeldovich}
Y. B. Zeldovich, ``Particle Creation in cosmology,'' 
Pisma Zh. Eksp. Teor. Fiz.  {\bf 12}, 443 (1970); 
Y.~B.~Zeldovich and A.~A.~Starobinsky, ``Particle production and vacuum polarization in an anisotropic gravitational field,'' 
Sov. Phys. JETP {\bf 34}, 1159 (1972). 

\bibitem{Parker:1974qw} 
  L.~Parker and S.~A.~Fulling,
  ``Adiabatic regularization of the energy momentum tensor of a quantized field in homogeneous spaces,''
  Phys.\ Rev.\ D {\bf 9}, 341 (1974).

\bibitem{mukhanov}
V. F. Mukhanov and S. Winitzki,
{\it Introduction to Quantum Effects in Gravity}, (Cambridge University Press, 2007).

\bibitem{hu}
E. A. Calzetta and B. L. Hu,
{\it Nonequilibrium Quantum Field Theory},
(Cambridge University Press, 2008).

\bibitem{Birrell:1982ix} 
  N.~D.~Birrell and P.~C.~W.~Davies,
  {\it Quantum Fields in Curved Space}, (Cambridge Univ Press, 1983).

\bibitem{mott} 
E.~Mottola,
``Particle Creation in de Sitter Space,''
Phys.\ Rev.\ D {\bf 31}, 754 (1985).

\bibitem{bousso}
R. Bousso, A. Maloney, and A. Strominger,
``Conformal vacua and entropy in de Sitter space'',
Phys. Rev. D {\bf 65}, 104039 (2002),
\hhref{hep-th/0112218}.

\bibitem{gds2010} 
E. Greenwood D. C Dai and D. Stojkovic,
``Time dependent fluctuations and particle production in cosmological de Sitter,''
Phys. Rev. Lett. B {\bf 692}, 226 (2010), 
\hhref{1008.0869}.
    
\bibitem{Polyakov:2007mm} 
  A.~M.~Polyakov,
  ``De Sitter space and eternity,''
  Nucl.\ Phys.\ B {\bf 797}, 199 (2008),
\hhref{0709.2899};
  ``Decay of Vacuum Energy,''
  Nucl.\ Phys.\ B {\bf 834}, 316 (2010),
\hhref{0912.5503}.

\bibitem{and}
P. R. Anderson and E. Mottola, ``On the Instability of Global de Sitter Space to Particle Creation," 
Phys. Rev. D {\bf 89}, 104038 (2014), 
\hhref{1310.0030};
``Quantum Vacuum Instability of `Eternal' de Sitter Space," 
Phys. Rev. D {\bf 89}, 104039 (2014), 
\hhref{1310.1963}.


\bibitem{gibbhawk}
G. W. Gibbons and S. W. Hawking, ``Cosmological Event Horizons, Thermodynamics, and Particle Creation'', 
Phys. Rev. D {\bf 15}, 2738 (1977).

\bibitem{ford}
L. H. Ford, ``Gravitational Particle Production and Inflation,'' 
Phys. Rev. D {\bf 35}, 2955 (1987).

\bibitem{Boyanovsky:1996rw} 
  D.~Boyanovsky, D.~Cormier, H.~J.~de Vega and R.~Holman,
  ``Out-of-equilibrium dynamics of an inflationary phase transition,''
  Phys.\ Rev.\ D {\bf 55}, 3373 (1997),
\hhref{hep-ph/9610396}.

\bibitem{vacha}
T.~Vachaspati, D.~Stojkovic and L.~Krauss,
``Observation of Incipient Black Holes and the Information Loss Problem,'' 
Phys. Rev. D, {\bf 76}, 024005 (2007), 
\hhref{gr-qc/0609024}; 
M.~Kolopanis and T.~Vachaspati,
``Quantum Excitations in Time-dependent Backgrounds,'' 
Phys. Rev. D {\bf 87}, 085041 (2013), 
\hhref{1302.1449}.

\bibitem{Brout:1995rd} 
  R.~Brout, S.~Massar, R.~Parentani and P.~Spindel,
  ``A Primer for black hole quantum physics,''
  Phys.\ Rept.\  {\bf 260}, 329 (1995),
\hhref{0710.4345}.
  
\bibitem{padma}
G.~Mahajan and T.~Padmanabhan,
``Particle Creation, Classicality, and Related Issues in Quantum Field Theory: I. Formalism and Toy Models,'' 
Gen. Rel. Grav. {\bf 40}, 661 (2007), 
\hhref{0708.1233}; 
``Particle Creation, Classicality, and Related Issues in Quantum Field Theory: II. Examples From Field Theory,'' 
Gen. Rel. Grav. {\bf 40}, 709 (2007), 
\hhref{0708.1237}.



\bibitem{unruh}
W. G. Unruh, ``Notes on blackhole evaporation,''
Phys. Rev. D {\bf 14}, 870 (1976).

\bibitem{Schutzhold:2006gj} 
  R.~Schutzhold, G.~Schaller and D.~Habs,
  ``Signatures of the Unruh effect from electrons accelerated by ultra-strong laser fields,''
  Phys.\ Rev.\ Lett.\  {\bf 97}, 121302 (2006),
  Erratum: [Phys.\ Rev.\ Lett.\  {\bf 97}, 139902 (2006)],
\hhref{quant-ph/0604065}.

\bibitem{gelisvenu}
F. Gelis and R. Venugopalan, ``Particle production in field theories coupled to strong external sources, I: Formalism and main Results,'' 
Nucl. Phys. A {\bf 776}, 135 (2006), 
\hhref{hep-ph/0601209}; 
``Particle production in field theories coupled to strong external sources, II: Generating Functions,'' 
Nucl. Phys. A {\bf 779}, 177 (2006), 
\hhref{hep-ph/0605246}.

\bibitem{kharzeevtuchin}
D.~Kharzeev and K.~Tuchin, ``From color glass condensate to quark gluon plasma through the event horizon,'' 
Nucl.\ Phys.\  A {\bf 753}, 316 (2005),
\hhref{hep-ph/0501234};
D.~Kharzeev, E.~Levin and K.~Tuchin, ``Multi-particle production and thermalization in high-energy QCD,'' 
Phys.\ Rev.\  C {\bf 75}, 044903 (2007),
\hhref{hep-ph/0602063}.

\bibitem{Gelis:2010nm} 
  F.~Gelis, E.~Iancu, J.~Jalilian-Marian and R.~Venugopalan,
  ``The Color Glass Condensate,''
  Ann.\ Rev.\ Nucl.\ Part.\ Sci.\  {\bf 60}, 463 (2010),
\hhref{1002.0333}


\bibitem{ruff}
R. Ruffini, G. Vereshchagin and S. Xue,
``Electron-positron pairs in physics and astrophysics: From heavy nuclei to black holes,''
J. Phys. Rep. {\bf 407}, 1 (2010),
\hhref{0812.3163}.


\bibitem{mourou}
 G.~Mourou, T.~Tajima and S.~Bulanov,
``Optics in the relativistic regime,''
Rev.\ Mod.\ Phys.\  {\bf 78}, 309 (2006).

\bibitem{mattias}
M.~Marklund and P.~Shukla,
``Nonlinear collective effects in photon photon and photon plasma
interactions,''
Rev.\ Mod.\ Phys.\  {\bf 78}, 591 (2006),
\hhref{hep-ph/0602123}.

\bibitem{dunne-eli}
G.~V.~Dunne,
``New Strong-Field QED Effects at ELI: Nonperturbative Vacuum Pair Production,''
Eur. Phys. J. D {\bf 55}, 327 (2009),
\hhref{0812.3163}.

\bibitem{DiPiazza:2011tq} 
  A.~Di Piazza, C.~Muller, K.~Z.~Hatsagortsyan and C.~H.~Keitel,
  ``Extremely high-intensity laser interactions with fundamental quantum systems,''
  Rev.\ Mod.\ Phys.\  {\bf 84}, 1177 (2012),
\hhref{111.3886}.



 \bibitem{oka}
T. Oka and H. Aoki,
``Nonequilibrium Quantum Breakdown in a Strongly Correlated Electron System'', \hhref{0803.0422v1}, in {\it 
Quantum and Semi-classical Percolation and Breakdown in Disordered Solids}, 
Lecture Notes in Physics, Vol. 762, A. K. Sen , K. K. Bardhan and B. K. Chakrabarti (Eds), (Springer, 2009).

\bibitem{Nation}
P.~D.~Nation, J.~R.~Johansson, M.~P.~Blencowe and F.~Nori,
``Colloquium: Stimulating uncertainty: Amplifying the quantum vacuum with superconducting circuits'',
Rev. Mod. Phys. 84, 1 (2012)
\href{1103.085}.

\bibitem{lzs}
  S.~N.~Shevchenko, S.~Ashhab and F.~Nori,
  ``Landau-Zener-Stuckelberg interferometry,''
  Phys.\ Rept.\  {\bf 492}, 1 (2010),
\hhref{0911.1917}.

\bibitem{Dodonov:2010zza} 
  V.~V.~Dodonov,
  ``Current status of the dynamical Casimir effect,''
  Phys.\ Scripta {\bf 82}, 038105 (2010).


\bibitem{nori}
C. M. Wilson, G. Johansson, A. Pourkabirian, M. Simoen,	J. R. Johansson, T. Duty,	F. Nori and P. Delsing,
``Observation of the dynamical Casimir effect in a superconducting circuit'',   Nature
    {\bf 479},   376   (2011). 

\bibitem{AkkermansDunne:2012}
E. Akkermans and G. V. Dunne, ``Ramsey Fringes and Time-domain Multiple-Slit Interference from Vacuum,'' 
Phys. Rev. Lett. {\bf 108}, 030401 (2012),
\hhref{1109.3489}.

\bibitem{reulet}
J. Gabelli and B. Reulet,
``Shaping a time-dependent excitation to minimize shot nise in a tunnel junction'',
Phys. Rev. B {\bf 87}, 075403 (2013).

\bibitem{brezin}
E. Brezin and C. Itzykson, ``Pair Production In Vacuum By An Alternating Field," 
Phys. Rev. D {\bf 2}, 1191 (1970).

\bibitem{popov}
V. S. Popov, ``Pair Production in a Variable External Field (Quasiclassical approximation)," 
Sov. Phys. JETP {\bf 34}, 709 (1972); 
``Pair production in a variable and homogeneous electric fields as an oscillator problem,"
Sov. Phys. JETP {\bf 35}, 659 (1972).

\bibitem{gavrilov}
V.~G.~Bagrov, D.~M.~Gitman, S.~P.~Gavrilov and S.~M.~Shvartsman,
  ``Creation Of Boson Pairs In A Vacuum,''
  Izv.\ Vuz.\ Fiz.\  {\bf 3}, 71 (1975);
  D.~Gitman and S.~Gavrilov,
  ``Quantum Processes In A Strong Electromagnetic Field. Creating Pairs",
  Izv.\ Vuz.\ Fiz.\  {\bf 1}, 94 (1977);
  S.~P.~Gavrilov and D.~M.~Gitman,
  ``Vacuum instability in external fields,''
  Phys. Rev. D{\bf 53}, 7162 (1996)
  \hhref{hep-th/9603152}.



\bibitem{KME:1998}
Y. Kluger, E. Mottola and J. Eisenberg, ``The quantum Vlasov equation and its Markov limit," 
Phys. Rev. D {\bf 58}, 125015 (1998), 
\hhref{hep-ph/9803372}.

\bibitem{Habib:1999cs} 
  S.~Habib, C.~Molina-Paris and E.~Mottola,
  ``Energy momentum tensor of particles created in an expanding universe,''
  Phys.\ Rev.\ D {\bf 61}, 024010 (2000),
\hhref{gr-qc/9906120}.

  
\bibitem{Rau:1995ea}
J.~Rau,
  ``Pair production in the quantum Boltzmann equation,''
  Phys.\ Rev.\  {\bf D50}, 6911 (1994)
  \hhref{hep-ph/9402256};
  J.~Rau and B.~M\"uller,
  ``From reversible quantum microdynamics to irreversible quantum transport,''
  Phys.\ Rept.\  {\bf 272}, 1-59 (1996)
  \hhref{nucl-th/9505009}.
  
\bibitem{schmidt}
  S.~A.~Smolyansky, G.~Ropke, S.~M.~Schmidt, D. Blaschke, V. D. Toneev, A. V. Prozorkevich,
  ``Dynamical derivation of a quantum kinetic equation for particle production in the Schwinger mechanism,''
  \hhref{hep-ph/9712377};
  S.~M.~Schmidt, D.~Blaschke, G.~Ropke, S. A. Smolyansky, A. V. Prozorkevich, V. D. Toneev, 
  ``A Quantum kinetic equation for particle production in the Schwinger mechanism,''
  Int.\ J.\ Mod.\ Phys.\  {\bf E7}, 709 (1998), 
  \hhref{hep-ph/9809227}.
  
\bibitem{Huet:2014mta} 
  A.~Huet, S.~P.~Kim and C.~Schubert,
  ``Vlasov equation for Schwinger pair production in a time-dependent electric field,''
  Phys.\ Rev.\ D {\bf 90}, no. 12, 125033 (2014),
\hhref{1411.3074}.

   \bibitem{kimpage}
 S.~P.~Kim and D.~Page,
  ``Schwinger pair production via instantons in a strong electric field,''
  Phys.\ Rev.\ D {\bf 65}, 105002 (2002)
  \hhref{hep-th/0005078},
  ``Schwinger pair production in electric and magnetic fields,''
 Phys.\ Rev.\  D {\bf 73}, 065020 (2006)
  \hhref{hep-th/0301132};
  ``Improved approximations for fermion pair production in inhomogeneous electric fields,''
  Phys.\ Rev.\  D {\bf 75}, 045013 (2007)
  \hhref{hep-th/0701047}.
    
\bibitem{Hebenstreit:2010vz} 
  F.~Hebenstreit, R.~Alkofer and H.~Gies,
  ``Schwinger pair production in space and time-dependent electric fields: Relating the Wigner formalism to quantum kinetic theory,''
  Phys.\ Rev.\ D {\bf 82}, 105026 (2010),
\hhref{1007.1099}.

\bibitem{Hebenstreit:2010cc} 
  F.~Hebenstreit, A.~Ilderton, M.~Marklund and J.~Zamanian,
  ``Strong field effects in laser pulses: the Wigner formalism,''
  Phys.\ Rev.\ D {\bf 83}, 065007 (2011),
\hhref{1011.1923}.




\bibitem{winitzki}
S. Winitzki, ``Cosmological particle production and the precision of the WKB approximation,'' 
Phys. Rev. D {72}, 104011 (2005), 
\hhref{gr-qc/0510001}.

\bibitem{Kim:2011jw} 
  S.~P.~Kim and C.~Schubert,
  ``Non-adiabatic Quantum Vlasov Equation for Schwinger Pair Production,''
  Phys.\ Rev.\ D {\bf 84}, 125028 (2011),
\hhref{1110.0900}. 
  
\bibitem{Zahn:2015awa} 
  J.~Zahn,
  ``The current density in quantum electrodynamics in time-dependent external potentials and the Schwinger effect,''
  J.\ Phys.\ A {\bf 48}, 475402 (2015),
\hhref{1501.06527}.

\bibitem{Gelis:2015kya} 
  F.~Gelis and N.~Tanji,
  ``Schwinger mechanism revisited,''
  Prog.\ Part.\ Nucl.\ Phys.\  {\bf 87}, 1 (2016),
\hhref{1510.05451}.

\bibitem{DabrowskiDunne:2014}
R.~Dabrowski and G.~V.~Dunne,
``Super-adiabatic Particle Number in Schwinger and de Sitter Particle Production,'' 
Phys. Rev. D {\bf 90}, 025021 (2014), 
\hhref{1405.0302}.

\bibitem{dingle}
R.~B.~Dingle, \emph{Asymptotic expansions: their derivation and interpretation}, (Academic Press, London,1973).


\bibitem{BerryAsymptotic}
M. V. Berry, ``Uniform asymptotic smoothing of Stoke's discontinuities,''
Proc. R. Soc. A {\bf 422}, 7 (1989);
``Waves near Stokes lines,'' 
Proc. R. Soc. A {\bf 427}, 265 (1990);
``Semiclassically weak reflections above analytic and non-analytic potential barriers''
J. Phys. A {\bf 15},  3693 (1982).

\bibitem{dumludunne}
C. K. Dumlu and G. V. Dunne, ``The Stokes Phenomenon and Schwinger Vacuum Pair Production in Time-Dependent Laser Pulses," 
Phys. Rev. Lett. {\bf 104}, 250402 (2010), 
\hhref{1004.2509}; 
``Interference Effects in Schwinger Vacuum Pair Production for Time-Dependent Laser Pulses," 
Phys. Rev. D {\bf 83}, 065028 (2011), 
\hhref{1102.2899}.

\bibitem{Fukushima:2014} 
K.~Fukushima,
``Spectral Representation of the Particle Production Out of Equilibrium - Schwinger Mechanism in Pulsed Electric Fields,'' 
New J. Phys. {\bf 16} (2014), 
\hhref{1402.3002}.

\bibitem{FukushimaHataya:2014} 
K.~Fukushima and T.~Hayata,
``Schwinger Mechanism with Stochastic Quantization,'' 
Phys. Lett. B {\bf 30} (2014),
\hhref{1403.4177}.


\bibitem{lim-berry}
R. Lim and M.V. Berry, ``Superadiabatic Tracking of Quantum Evolution'',  
J.Phys.A {\bf 24}, 3255 (1991);
R. Lim, ``Overlapping Stokes smoothings in adiabatic quantum transitions'',
J. Phys. A: Math. Gen. {\bf 26}, 7615 (1993).

\bibitem{BB}
M.~V.~Berry, ``Transitionless quantum driving,''
J. Phys. A: Math. Theor. {\bf 42}, 365303 (2009).

\bibitem{expberry}
G. Tayebirad, A. Zenesini, D. Ciampini, R.~Mannella, O. Morsch, E. Arimondo, N. L\"orch and S. Wimberger,
``Time-resolved measurement of Laundau-Zener tunneling in different bases,'' 
Phys. Rev. A {\bf 82}, 013633 (2010), [Erratum, Phys. Rev. A {\bf 82}, 069904 (2010)].

\bibitem{demirice}
M. Demirplak and S. A. Rice, ``Adiabatic Population Transfer with Control Fields,'' 
Phys. Chem. A {\bf 107}, 9937 (2003).



\bibitem{delCampo}
A. del Campo, ``Shortcuts to adiabaticity by counter-diabatic driving,''
Phys. Rev. Lett. {\bf 111}, 100502 (2013), 
\hhref{1306.0410}.

\bibitem{jarz-shortcut}
C. Jarzynski, ``Generating shortcuts to adiabaticity in quantum and classical dynamics,'' 
Phys. Rev. A {\bf 88}, 040101(R) (2013),
\hhref{1305.4967}.


\bibitem{CDexp1}
M. G. Bason, M. Viteau, N. Malossi, P. Huillery, E. Arimondo, D. Ciampini, R. Fazio, V. Giovannetti, R. Mannella and O. Morsch, 
``High-fidelity quantum driving,'' 
Nat. Phys. {\bf 8}, 147 (2012).

\bibitem{CDexp2}
J. Zhang, J. H. Shim, I. Niemeyer, T. Taniguchi, T. Teraji, H. Abe, S. Onoda, T. Yamamoto, T. Ohshima, J. Isoya, and D. Suter, 
``Experimental Implementation of Assisted Quantum Adiabatic Passage in a Single Spin,''
Phys. Rev. Lett. {\bf 110}, 240501 (2013).

\bibitem{steen}
A. Steen, ``Om Formen for Integralet af den lineaere Differentialligning af anden Orden", 
Overs. over d. K. Danske Vidensk. Selsk. Forh. (1874), 1-12; 
R. Redheffer and I. Redheffer, ``Steen's 1874 paper: historical survey and translation,''
Appl. Anal. Discrete Math. {\bf 2} (2008), 146-157.

\bibitem{erm}
V. P. Ermakov, ``Second-order Differential Equations: Conditions of Complete Integrability,'' 
Univ. Izv. Kiev {\bf 20}, 1 (1880).

\bibitem{milne}
W.~E.~Milne, ``The Numerical Determination of Characteristic Numbers,''
Phys. Rev. {\bf 35}, 863 (1930).

\bibitem{pinney}
E. Pinney, ``The Nonlinear Differential Equation $y''+p(x) y + c y^{-3}=0$,''
Proc. Am. Math. Soc. {\bf 1}, 681 (1950).

\bibitem{husimi}
K. Husimi, ``Miscellanea in elementary quantum mechanics: I-II'', Prog. Theor. Phys. {\bf 9}, 238-244 (1953); 
Prog. Theor. Phys. {\bf 9}, 381-402 (1953).

\bibitem{DittrichReuter}
W.~Dittrich and M.~Reuter,
\emph{Classical and Quantum Dynamics}, (Springer, 2001).

\bibitem{LewisInvariant1}
H.~R.~Lewis, ``Classical and Quantum Systems with Time-Dependent Harmonic-Oscillator-Type Hamiltonians,''
Phys. Rev. Lett. {\bf 18} 518 (1967); 
``Class of Exact Invariants for Classical and Quantum Time-Dependent Harmonic Oscillators,''
J. Math. Phys. {\bf 9}, 1976 (1968);
H.~R.~Lewis and W.~B.~Riesenfeld, ``An Exact Quantum Theory of the Time-Dependent Harmonic Oscillator and of a Charged Particle in a Time-Dependent Electromagnetic Field,'' 
J. Math. Phys. {\bf 10}, 1458 (1969).


\bibitem{gelfand}
I.~M.~Gel'fand and L.~A.~Dikii, ``Asymptotic Behavior of the Resolvent of Sturm-Liouville Equations and the Algebra of the Korteweg-De Vries Equations,''
Russ. Math. Surv. {\bf 30}, 5 (1975);
``The Resolvent and Hamiltonian Systems,''
Func. Anal. App. {\bf {11}}, 2 (1977), 93-105.

\bibitem{Balantekin:1990aa} 
  A.~B.~Balantekin, J.~E.~Seger and S.~H.~Fricke,
  ``Dynamical effects in pair production by electric fields,''
  Int.\ J.\ Mod.\ Phys.\ A {\bf 6}, 695 (1991).


\bibitem{dunnehall}
G.~V.~Dunne and T.~Hall, ``On the QED Effective Action in Time-Dependent Electric Backgrounds,''
Phys. Rev. D {\bf 58}, 105022 (1998), 
\hhref{hep-th/9807031}.


\bibitem{budden}
K.~G.~Budden, \emph{Radio Waves in the Ionosphere: The Mathematical Theory of the Reflection of Radio Waves from Stratified Ionised Layers}, 
(Cambridge Univ. Press, 1961).

\bibitem{berry-mount}
M.~V.~Berry and K.~E.~Mount, ``Semiclassical approximations in wave mechanics,''
Rept. Prog. Phys. {\bf 35}, 315 (1972)

\bibitem{Narozhnyi:1970uv} 
  N.~B.~Narozhnyi and A.~I.~Nikishov,
  ``The Simplest processes in the pair creating electric field,''
  Yad.\ Fiz.\  {\bf 11}, 1072 (1970)
  [Sov.\ J.\ Nucl.\ Phys.\  {\bf 11}, 596 (1970)].


\end{thebibliography}
\end{document}